# Generic approach for the modeling of liquefied thermochemical products and biomass heat of formation. Case study: HTL biocrude, Pyrolysis oil and assessment of energy requirements

E.M. Lozano, T.H. Pedersen, L.A. Rosendahl [1]

Department of Energy Technology, Aalborg University, Pontoppidanstræde 111, 9220 Aalborg Øst, Denmark

**Abstract:** A generic approach is proposed for the estimation of advanced biocrudes properties from liquefied biomass and the enthalpy of formation of biomass feedstocks applicable to the modeling of biomass conversion processes where the exact stoichiometry and kinetics are unknown,such as pyrolysis, solvolysis and hydrothermal liquefaction (HTL) . The enthalpy of formation of the biomass is estimated through a direct correlation based on ultimate and proximate analysis, whose parameters can easily be fitted with experimental data available from sources such as the Phillys database for different biomass types and implemented in process simulators such as Aspen Plus®. For the biocrude modeling, a multi-objective optimization model is proposed that refines the selection of model compounds to match measured bulk thermochemical and physical properties. Parameter fitting and multi-objective optimization were both performed in Matlab® and the codes are available in the supplementary material. As a case study, the optimization model was applied to two different oils obtained via HTL and pyrolysis of woody biomass. The first case was further studied in order to estimate the impact of the proposed models in the energy requirements. The process was implemented in Aspen Plus® but the methodology is applicable to other simulation tools. The results show that the methods for estimating enthalpies of formation have a high impact on the energy balance and consequently the models developed allow a more accurate estimation of the energy requirement in the reactor. This is a key element in making accurate heat integration and techno-economic analyses of thermochemical conversion processes.

## 1. INTRODUCTION

Thermochemical conversion of biomass comprises different processes by which heat is used to transform biomass material into gas, liquid and/or solid fuels. Due to their feedstock flexibility and the possibility to obtain a wide range of products, thermochemical routes have gained increasing interest in recent years, motivated by the need to develop more sustainable, carbon neutral processes and products in the transition from linear to circular economies. Gasification, pyrolysis, carbonization, torrefaction and hydrothermal liquefaction (HTL) are examples of this.

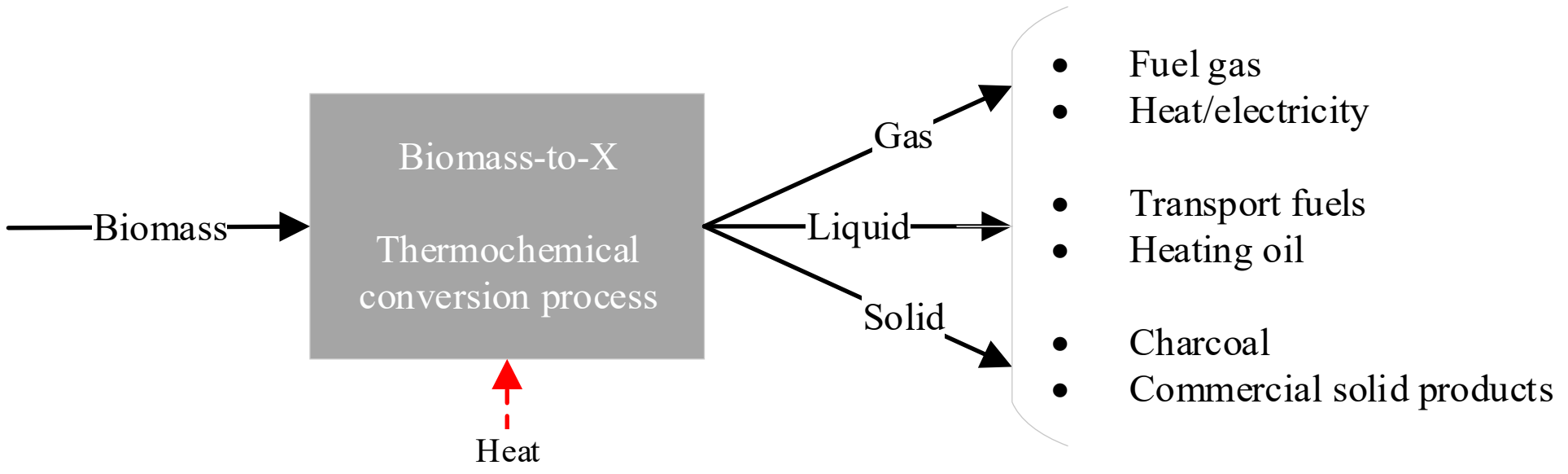

*Figure 1 Possible products of thermochemical conversion of biomass*

Compared to the biochemical routes, thermochemical processes are performed at relatively high temperatures and/or pressures and consequently, efficient heat integration is crucial to favor the economics of the process. From basic thermodynamics, the energy balance at reaction conditions (Hess's law) can be expressed as:

$$Q = \Delta H_{rxn}^{T,P} = \sum_n H_{f,prod}^{T,P} - \sum_n H_{f,react}^{T,P} \quad [J] \tag{1}$$

Where $H_{f,i}°$ represents the total enthalpy in [J] for each reactant and product. For solid and liquid components -such as biomass and biocrude oils- the effect of pressure in the enthalpy change is commonly neglected and the total enthalpy at reaction conditions is given by the summation of the formation enthalpy and sensible enthalpy:

$$h_{f,i}(T) = \Delta h_{f,i}° + \int_{T_0}^{T} C_{p,i}\, dT \quad [J/kg] \tag{2}$$

---

[1] Corresponding author. E-mail address: lar@et.aau.dk (L.A. Rosendahl).

Where $\Delta h_{f,i}°$ is the specific standard enthalpy of formation of component i in [J/kg], $C_{p,i}$ is the specific heat capacity in [J/kgK], T is the reaction temperature in [K], and $T_0$ is the standard temperature (298.15 K).

Even though the standard enthalpy of formation is commonly reported in literature for pure components and the calculation is straightforward for mixtures with well-defined composition, this is non-trivial in the case of complex streams comprising multiple components such as biomass and biocrude oils.

The prevalent method that has been reported in literature to estimate heat of formation of biomass is from its higher heating value (HHV), or lower heating value (LHV) in some cases. This approach is based on the combustion reaction (eq. 3) and has been widely used in the modeling of different processes such as torrefaction (Bates & Ghoniem, 2013), gasification and pyrolysis (La Villetta, Costa, & Massarottia, 2017) (Baratieri, Baggio, Fiori, & Grigiante, 2008). The complete combustion of biomass (represented as $C_\varphi H_\alpha O_\beta N_\gamma S_\delta$ in eq. 3) can be expressed as:

$$C_\varphi H_\alpha O_\beta N_\gamma S_\delta + \left(\varphi + \frac{\alpha}{4} - \frac{\beta}{2} + \gamma + \delta\right) O_2 \rightarrow \varphi CO_2 + \frac{\alpha}{2} H_2O + \gamma NO_2 + \delta SO_2 \qquad (3)$$

Applying the energy balance (eq. 1):

$$HHV = -\Delta H_{rxn} = \Delta h^0_{f,biomass} + \left(\varphi + \frac{\alpha}{4} - \frac{\beta}{2} + \gamma + \delta\right) \Delta_f h^0_{O_2} - \varphi \Delta_f h^0_{CO_2} - \frac{\alpha}{2} \Delta_f h^0_{H_2O} - \delta \Delta_f h^0_{SO_2} - \gamma \Delta_f h^0_{NO_2} \quad [J] \qquad (4)$$

Eq. 4 can be rewritten in terms of the mass fractions of C,H,O,N,S in the biomass ($w_i$) obtained experimentally from ultimate analysis and the known standard enthalpies of formation of the combustion products (water in liquid phase when the HHV is used). The composition, HHV and heat of formation must be in the same basis (dry (d) or dry ash-free (daf) basis). Solving for $\Delta_f h^{0,daf}_{biomass}$:

$$\varphi = \frac{w_C^{daf}}{MW_C}; \alpha = \frac{w_H^{daf}}{MW_H}; \beta = \frac{w_O^{daf}}{MW_O}; \gamma = \frac{w_N^{daf}}{MW_N}; \delta = \frac{w_S^{daf}}{MW_S} \quad \left[\frac{mol}{g}\right] \qquad (5)$$

$$\Delta h^{0,daf}_{f,biomass} = HHV^{daf} - \left(3{,}278 * 10^5\, w_C^{daf} + 1{,}418 * 10^6 w_H^{daf} + 9{,}264 * 10^4\, w_S^{daf} - 2{,}418 * 10^4\, w_N^{daf}\right) * 10^2 \quad \left[\frac{J}{kg}\right] \qquad (6)$$

The heat of formation can be estimated from experimental results of the HHV or using estimations from empirical correlations. Different correlations have been developed and are reported in literature for various types of biomass. Extensive reviews can be found for example in (Peduzzi, Boissonnet, & Maréchal, 2016) (Channiwala & Parikh, 2002) (A. Friedl, 2005) (Yin, 2011) in which existing correlations are listed with mean absolute errors in the range from 1.5 to 7% approximately. Even though the HHV correlations available in literature are relatively accurate, their use for estimating the heat of formation can induce a considerable error, due to the difference in magnitude of the heat of formation relative to the HHV that gives this term a high weight in eq. 6.

This has been mentioned as a limitation for the modeling of non-conventional compounds, since an error of 1% in the estimation of the HHV can produce up to 50% error when it is used to calculate the heat of formation (Aspen Technology Inc., 2006). A direct correlation for the estimation of the heat of formation from ultimate and proximate analysis is also available which was developed using data from the Penn State Data Base (Penn Sate University Libraries, 2019) and can be selected as an alternative to the HHV route; nevertheless, its use for biomass has not been verified in the literature although it is extensively used.

Alternative approaches have been used to estimate the heat of formation of compounds based on their molecular structure such as group contribution models, force fields models and quantitative structure-property relationship (QSPR) models (Villeda, Dahmen, Hechinger, & Voll, 2015). In the group contribution models, molecules are composed of structural units or groups that make a specific numerical contribution in the estimation of a macroscopic property. A more detailed discussion of the estimation of the enthalpy of formation from group additivities can be found in (Benson, 1976). In force field models, also referred to as molecular mechanics methods, molecules are described by a ball and spring system with atoms and bonds having different sizes, lengths and stiffness that can undergo rearrangement in the course of a chemical reaction. Thus, the enthalpy of formation of a compound can be derived from reaction enthalpies considering the elements involved (Villeda, Dahmen, Hechinger, & Voll, 2015) (Jensen, 2016). In the QSPR models, mathematical methods are used to find a relation between a macroscopic property of interest (e.g. heat of formation) and a variety of different attributes called "molecular descriptors" derived from the molecular structure (Vatani, Mehrpooya, & Gharagheizi, 2007) (Teixeira, Leal, & Falcao, 2013).

A different method was also reported by Vasiliu et al. who used advanced computational chemistry methods (G3MP2) based on the Gaussian-2 procedure (G2) to compute enthalpies of formation based on atomization energies for different biomass-derived compounds such as glucose, HFM, sorbitol, succinic acid, 2,5-furandicarboxylic acid (FDCA), 3-hydroxypropionic acid (3-HPA), among others, which have been identified as important biomass building blocks (Vasiliu, Guynn, & Dixon, 2011) (Vasiliu, Jones, Guynn, & Dixon, 2012).

However, the methods previously described are applicable to identifiable molecules for which their implementation in the prediction of the enthalpy of formation of the complex structure of biomass is not easily found in literature.

The enthalpy of formation of the liquid biofuels can be also derived following the HHV approach and, as for biomass, several empirical correlations are available in literature (Beckman, et al., 1990) (Channiwala & Parikh, 2002) (Fassinou, Steene, Toure, & Martin, 2011).

Among other methods in literature, Yang, et al. (Yang, et al., 2013) derived a correlation for the estimation of the enthalpy of formation of pyrolysis bio-oil as a function of H/C and O/C ratios. In the context of process design, however, the modeling of the produced biofuels goes beyond an accurate estimation of the heat of formation, due to the subsequent processing steps that normally involve separation, reaction, upgrading, etc. In many cases, this requires identifiable compounds to be modeled.

Following from this, model compound analysis has been a common approach to model biocrude-oils based on the selection of representative model compounds by the results of experimental characterization techniques such as GC-MS. Once the composition is defined, the enthalpy of formation as well as other physical/chemical properties can be computed. However, this method poses the difficulty that just the volatile fraction of the biocrude is analyzed, leaving a major fraction unknown and the results therefore severely error prone (Pedersen, Jensen, Sandström, & Rosendahl, 2017). Further, even within this analyzed fraction, the amount of identified components is enormous, requiring significant time for the analysis of the results. Previous studies of PNNL and VTT report the development and refining of the model compound lists for HTL and fast pyrolysis bio-oil, however, the is not further information on the methods used (Tews, et al., 2014). In other cases, the selection is based merely on the results of the GC-MS spectrogram without calibration and matching the properties of the biocrude, which can deviate significantly from the expected by experiments.

Other modeling strategies that have been proposed in order to characterize the produced components deal with the modeling of the reaction kinetics and aim to provide a more fundamental approach. Sharifzadeh and Shah (Sharifzadeh, Richard, & Shah, 2017) developed a model that uses a reaction network superstructure to classify the products of the deoxygenation of a pyrolysis-oil based on the –OH groups present. This work also summarizes the main limitations in the establishment of reaction pathways based on the review of previous kinetic studies in the field of hydrothermal upgrading, which are in general applicable to direct thermochemical liquefaction processes such as pyrolysis and HTL (Sheehan & Savage, 2017) (Valdez, Tocco, & Savage, 2014):

- Reactions are highly interactive and exhibit nonlinear behaviors
- Diverse and numerous components in the oils hinder the development of kinetic models based on actual species
- Existing kinetic models are often based on lumped modelling of the involved phases or conversion of the initial biomass but do not provide any insight into the identity of products or their *physical and chemical properties*

To the best of our knowledge, alternative methods for modeling biomass and biocrude-oil properties have not been reported. Based on this, the aim of this paper is: 1) to examine the accuracy of existing HHV correlations to estimate the heat of formation of biomass and propose and demonstrate an alternative correlation for woody biomass 2) propose and demonstrate an alternative, generic approach for the modeling of biocrude oils and implement it for two cases -HTL biocrude and pyrolysis oil- and 3) to evaluate the impact of the proposed methods in the estimation of the energy demand in the case of HTL of woody biomass. For the biomass feedstock, the aim is to compare different approaches for estimating the heat of formation and to verify the use of the direct correlation proposed in this work as an alternative and more accurate route. For the biocrude-oils, the focus is on the model compound approach from experimental results instead of from predictive kinetics/equilibrium models. The purpose is to develop a biocrude model that refines the selection of model compounds from experiments in order to match measurable properties and, at the same time, maintaining elemental and energy balances across the reactions. The methodology in this work is developed for woody biomass but can be easily applied to other biomass types.

## 2. METHODOLOGY

### 2.1 *Enthalpy of formation of biomass:*

A sample of 30 different types of wood was selected from Phillys database (ECN-TNO, 2019), consisting of experimental proximate and ultimate analyses and HHV (list of samples available in the supplementary material). For each sample, the HHV was estimated from the correlations in Table 1 (eq.7-12) and the heat of formation was obtained by inserting the estimated HHV in eq. 6.

*Table 1 Empirical correlations for the estimation of the HHV (Rönsch & Wagner, 2012) (Aspen Technology Inc., 2006)*

| Author | Empirical correlation for HHV | No. |
|---|---|---|
| Boie (BOIE) | $HHV^{daf} = \left(\left(a_1 w_C^{daf} + a_2 w_H^{daf} + a_3\, w_S^{daf} + a_4 w_O^{daf} + a_5 w_N^{daf}\right) * 10^2 + a_6\right) * 2326 \ \left[\frac{J}{kg}\right]$ | (7) |
| Dulong (DLNG) | $HHV^{daf} = \left(\left(a_1 w_C^{daf} + a_2 w_H^{daf} + a_3\, w_S^{daf} + a_4 w_O^{daf} + a_5 w_N^{daf}\right) * 10^2 + a_5\right) * 2326 \ \left[\frac{J}{kg}\right]$ | (8) |
| Grummel and Davis (GMLD) | $HHV^{daf} = \left(a_1 w_C^{daf} + w_H^{daf} + a_3\, w_S^{daf} + a_4 w_O^{daf}\left(\frac{a_2 w_H^{daf}}{1 - w_A^{d}} + a_5\right) * 10^2\right) * 2326 \ \left[\frac{J}{kg}\right]$ | (9) |
| Mott and Spooner (MTSP) | $HHV^{daf} = \left[\left(a_1 w_C^{daf} + a_2 w_H^{daf} + a_3\, w_S^{daf} - \left(a_6 + \frac{a_5 w_O^{daf}}{1 - w_A^{d}}\right) w_O^{daf}\right) * 10^2 + a_7\right] * 2326 \ \left[\frac{J}{kg}\right]$ | (10) |
| IGT | $HHV^{daf} = \left(\left(a_1 w_C^{d} + a_2 w_H^{d} + a_3\, w_S^{d} + a_4 w_A^{d}\right) * 10^2 + a_5\right) * 2326 \ \left[\frac{J}{kg}\right]$ | (11) |
| Revised IGT (IGT2) | $HHV^{daf} = \left(\left(a_1 w_C^{d} + a_2 w_H^{d} + a_3\, w_S^{d} + a_4 w_A^{d} + a_5\left(w_O^{d} + w_N^{d}\right)\right) * 10^2\right) * 2326 \ \left[\frac{J}{kg}\right]$ | (12) |

The heat of formation was also estimated from the direct correlation (DC) (eq. 13) and a modified version for biomass, developed in this work, referred to from here as *Lozano* correlation (LZN) (eq. 14).

Direct correlation (DC)

$$\Delta h_f^{0,daf} = \left[\left(a_1 w_C^{dm} + a_2 w_H^{dm} + a_3 w_H^d + a_4 w_{S_p}^d + a_5 w_{Ss}^d\right) 10^2 + a_6 R_o + \left(a_7\left(w_C^d - w_{FC}^d\right) + a_8 w_{VM}^d\right) 10^2 + \left(a_9\left(w_C^{dm}\right)^2 + a_{10}\left(w_S^{dm}\right)^2 + a_{11}\left(w_C^d - w_{FC}^d\right)^2 + a_{12}\left(w_{VM}^d\right)^2\right) 10^4 + a_{13}(R_o)^2 + a_{14}\left(w_{VM}^d\right)\left(w_C^d - w_{FC}^d\right) 10^4 + a_{15}\right] * \frac{2326}{(1-w_A^d)} \quad \left[\frac{J}{kg}\right] \tag{13}$$

Lozano correlation (LZN)

$$\Delta h_f^{0,daf} = \left[\left(a_1 w_C^{dm} + a_2 w_H^{dm} + a_3 w_H^d\right) 10^2 + \left(a_4\left(w_C^d - w_{FC}^d\right) + a_5 w_{VM}^d\right) 10^2 + \left(a_6\left(w_C^{dm}\right)^2 + a_7\left(w_S^{dm}\right)^2 + a_8\left(w_C^d - w_{FC}^d\right)^2 + a_9\left(w_{VM}^d\right)^2\right) 10^4 + a_{10}\left(w_{VM}^d\right)\left(w_C^d - w_{FC}^d\right) 10^4 + a_{11}\right] * \frac{2326}{(1-w_A^d)} \quad \left[\frac{J}{kg}\right] \tag{14}$$

Where $a_i$ are the parameters for each correlation summarized in Table 2 , and $w_i^{dm}$ and $w_i^d$ are the reported weight percentages in dry ash-free and dry basis respectively of carbon (C), hydrogen (H), sulfur (S), oxygen (O), fixed carbon (FC), volatile matter (VM) and (A) ash. The parameter Ro in eq. 13 represents the mean-maximum vitrinite reflectance for coals, not included for the calculations of wood.

*Table 2 Parameters of HHV correlations and direct correlation*

| Parameter (a) | BOIE | DLNG | GMLD | MTSP | IGT | IGT2 | DC |
|---|---|---|---|---|---|---|---|
| 1 | 151.2 | 145.44 | 0.3333 | 144.54 | 198.11 | 146.58 | 1810.12 |
| 2 | 499.77 | 620.28 | 654.3 | 610.2 | 620.31 | 568.78 | -502.22 |
| 3 | 45 | 40.5 | 0.125 | 40.3 | 80.93 | 29.4 | 329.109 |
| 4 | -47.7 | -77.54 | -0.125 | 62.45 | 44.95 | -6.58 | 121.766 |
| 5 | 27 | -16 | 424.62 | 30.96 | -5153 | -51.53 | -542.39 |
| 6 | -189 | | -2 | 35.88 | | | 1601.57 |
| 7 | | | | -47 | | | 424.25 |
| 8 | | | | | | | -525.2 |
| 9 | | | | | | | -11.481 |
| 10 | | | | | | | 31.585 |
| 11 | | | | | | | 13.5256 |
| 12 | | | | | | | 11.5 |
| 13 | | | | | | | -685.85 |
| 14 | | | | | | | -22.494 |
| 15 | | | | | | | -64836 |

The parameters of eq. 14 were fitted in Matlab® using the *lsqnonlin* solver for non-linear least square problems to minimize an error function. This function computes the error between experimental and predicted HHVs, estimated by the adjusted correlation (LZN) from eq.6. The Matlab® code is provided in the supplementary material.

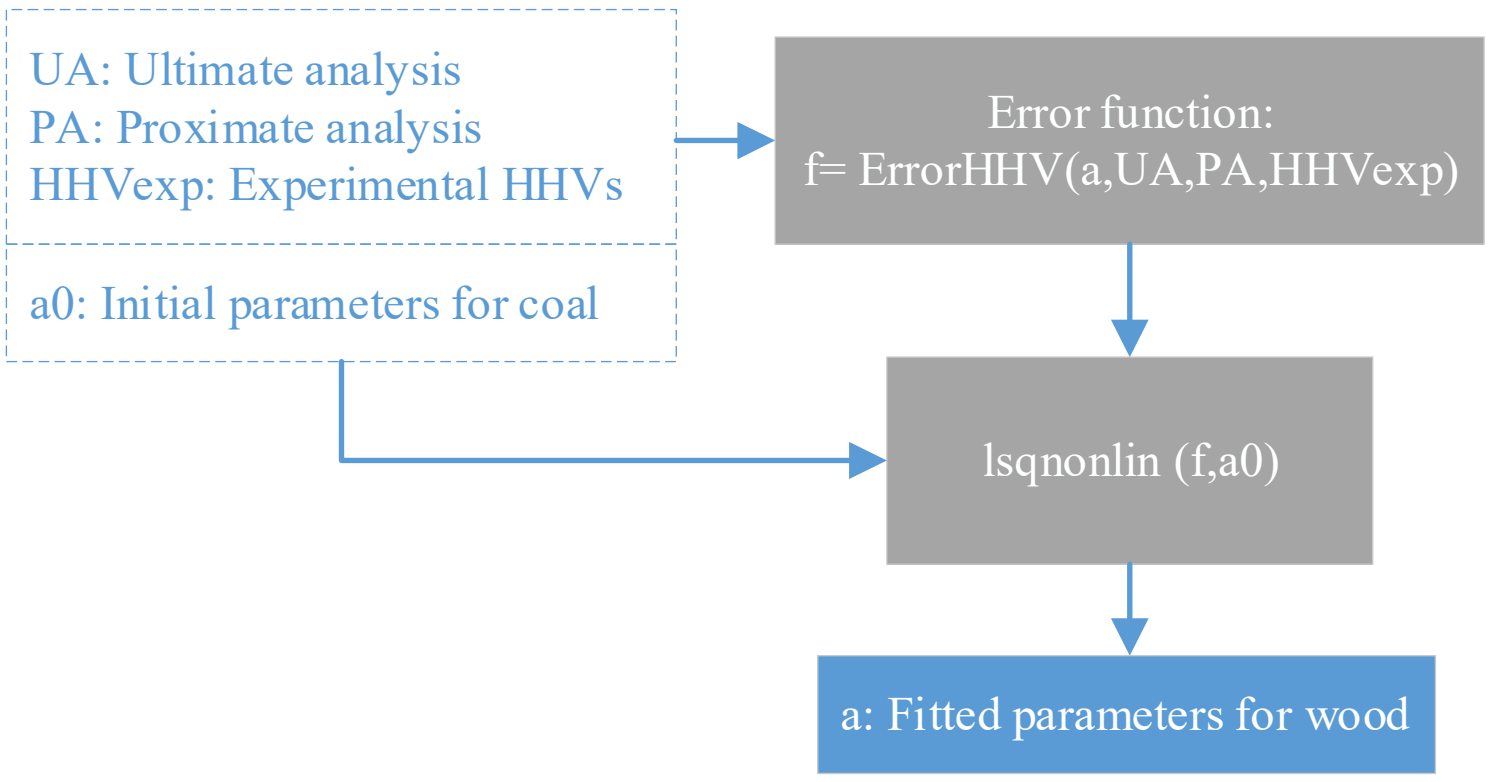

*Figure 2 Flow diagram of parameter estimation of adjusted direct correlation*

The results of the correlation of this work were compared with the correlations in Table 1. The performance of the adjusted direct correlation was evaluated through leave-1-out cross validation (jackknife procedure) in which one observation is left out of the calibration data set, the model is recalibrated, and the observation that was left out is then predicted. The root mean square error (RMSE) was computed for the full model and the leave-1-out validation.

## 2.2 *Biocrude model*

A biocrude model was developed based on multi-objective optimization to determine an optimum distribution of model compounds that satisfies simultaneously different objective functions. The objective functions formulated (eq. 15-20) correspond to the error between the experimental biocrude properties and the estimated ones based on the properties of individual model compounds:

$$\Delta h_{f,exp} - \sum_{i=1}^{n} x_i \, \Delta h_{f,i} = 0 \tag{15}$$

$$C(wt\%)_{exp} - \sum_{i=1}^{n} x_i \, C_i(wt\%) = 0 \tag{16}$$

$$H(wt\%)_{exp} - \sum_{i=1}^{n} x_i \, H_i(wt\%) = 0 \tag{17}$$

$$O\,(wt\%)_{exp} - \sum_{i=1}^{n} x_i \, O_i(wt\%) = 0 \tag{18}$$

$$\rho_{exp} - \sum_{i=1}^{n} x_i \, \rho_{,i} = 0 \tag{19}$$

$$1 - \sum_{i=1}^{n} x_{,i} = 0 \tag{20}$$

Where $\Delta h_{f,}$, C/H/O $(wt\%)$, $\rho$ and $Cp$ represent the specific heat of formation, elemental composition, density and heat capacity respectively; the subscript exp indicates the biocrude property from experimental results and the subscript i the property of the individual model compound, and $x_i$ represent the mass fractions of each model compound.

The multi-objective optimization was performed using the goal attainment method. In this algorithm, a vector of objective functions ($F_i(x)$) and designed goals ($F_i^*$, =0 in this case) are defined, and the relative degree of under- or over-achievement of the goals is controlled by weighting coefficients ($w_i$) -a hard constraint is introduced when a weighting coefficient is set to zero-. The objective is to minimize the parameter $\gamma$ *(De Schutter, 2018)*:

$$F_i(x) - w_i\gamma \leq F_i^* \tag{21}$$

The model was developed in Matlab® using the function *fgoalattain* and the code is provided in the supplementary material.

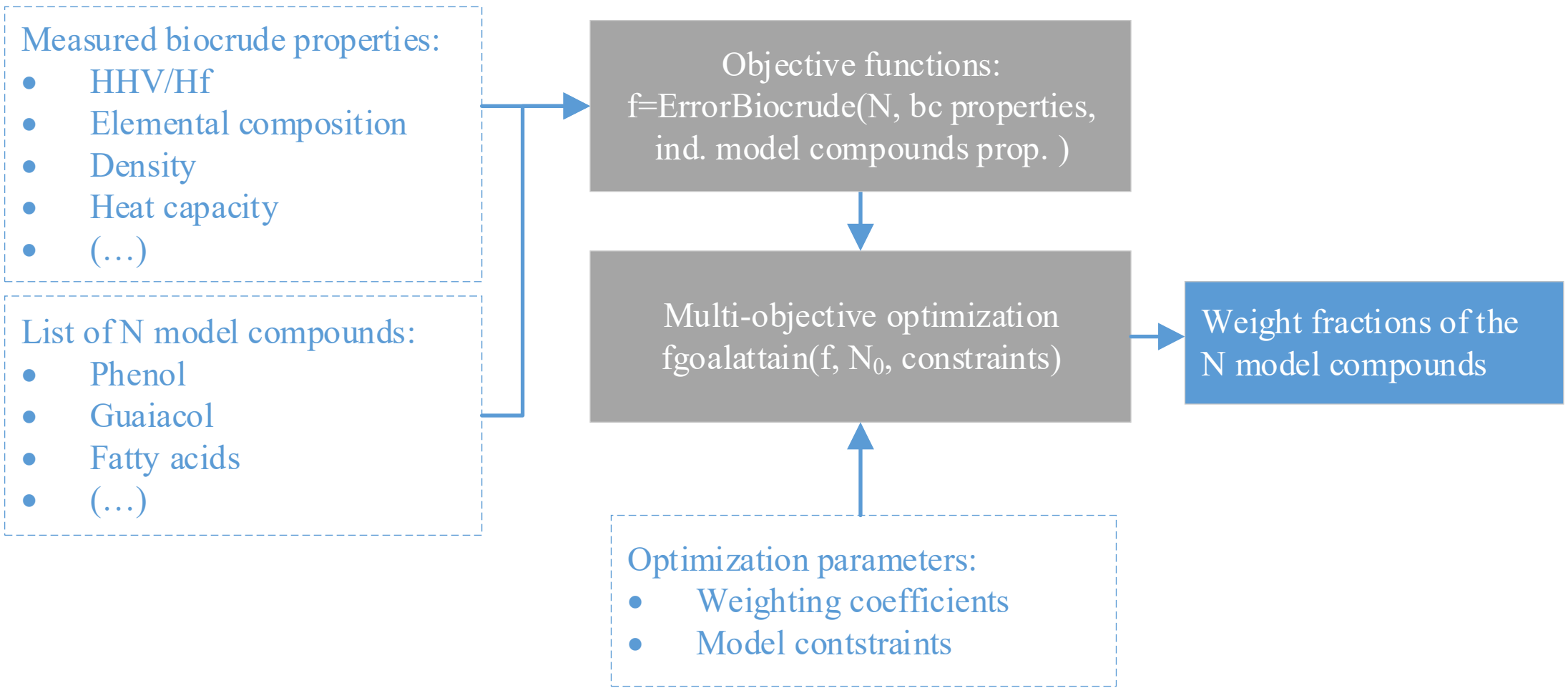

*Figure 3 Flow diagram for estimating biocrude properties*

The biocrude model was implemented for two cases: HTL biocrude from pinewood and pyrolysis oil from spruce wood. In the first case, available experimental results in-house of the characterization of a biocrude sample were used; in the second case, the data was obtained from experimental results reported in literature (Demirbas F. , 2010) (Demirbas A. , Mechanisms of liquefaction and pyrolysis reactions of biomass, 2000). Table 3 and Table 4 show the normalized composition of model compounds for each case. The mass percentage, HHV, elemental composition and density were determined experimentally (Table 5). The properties of the model compounds were obtained from Aspen Plus® library using the SRK property method. Furthermore, taking into account that the results of the GC-MS are only representative of the volatile fraction of the biocrude oils (the temperature ramp is typically set until 280º C), additional compounds are needed in the lists in order to represent the heavy fraction. In a previous study on HTL biocrude, Pedersen and collaborators (Pedersen, Jensen, Sandström, & Rosendahl, 2017) reported the full characterization of compounds as well as the HHV, elemental composition and true boiling point of different fractions obtained through fractional distillation. The results reported for the heavy fractions (HVF) whose normal boiling point is higher than 300º C are used in this study to complement the model compounds and are also presented in Table 5. These were also used for the pyrolysis oil case since similar data was not easily found in literature of the subject.

*Table 3 Model compounds identified in GC-MS of HTL biocrude sample.*

| | Model compound | wt.% (db) | | Model compound | wt.% (db) |
|---|---|---|---|---|---|
| 1 | Ethanol | 0.40 | 12 | 2,3-Dimethylphenol | 5.45 |
| 2 | Methyl-ethyl-ketone | 0.43 | 13 | Phenol, 2,6-dimethyl | 2.53 |
| 3 | Methyl-isopropyl-ketone | 0.35 | 14 | Caproic acid | 0.99 |
| 4 | Methyl-n-propyl-ketone | 0.42 | 15 | Heptanoic acid | 1.80 |
| 5 | 2-Cyclopenten-1-one, 2,3-dimethyl- | 2.23 | 16 | Octanoic acid | 2.15 |
| 6 | Phenol | 1.82 | 17 | 4-methyl catechol | 14.21 |
| 7 | 2-Cyclopenten-1-one, 3,4,4-trimethyl- | 2.51 | 18 | 2-Hydroxyphenethyl alcohol (isomer) | 18.02 |
| 8 | 2-Cyclopenten-1-one, 2,3,4,5-tetramethyl- | 1.03 | 19 | Benzeneacetic acid, 3-hydroxy | 5.80 |
| 9 | 2-Methylphenol | 2.01 | 20 | Palmitic acid | 9.71 |
| 10 | Phenol, 4-methyl | 4.52 | 21 | Myristic acid | 4.70 |
| 11 | 2,5-Dimethylphenol | 2.11 | 22 | Octadecanoic acid | 16.83 |

*Table 4 Model compounds in pyrolysis oil (Demirbas A. , Mechanisms of liquefaction and pyrolysis reactions of biomass, 2000)*

| | Model compound | wt.% (db) | | Model compound | wt.% (db) |
|---|---|---|---|---|---|
| 1 | Acetic-acid | 38.92 | 15 | Valeric-acid | 1.25 |
| 2 | 1-hydroxy-2-propanone | 16.26 | 16 | Isovaleric-acid | 1.25 |
| 3 | Methanol | 9.53 | 17 | 5-methylfurfural | 0.77 |
| 4 | Furfural | 4.85 | 18 | Valerolactone | 0.77 |
| 5 | 2,6-dimethoxyphenol | 4.57 | 19 | n-butyric-acid | 0.77 |
| 6 | Levoglucosan | 4.38 | 20 | Isobutyric-acid | 0.77 |
| 7 | Guaiacol | 1.90 | 21 | n-propionaldehyde | 0.77 |
| 8 | Crotonic-acid | 1.55 | 22 | Acrylic-acid | 0.77 |
| 9 | Formic-acid | 1.53 | 23 | Methyl-ethyl-ketone | 0.77 |
| 10 | Butyrolactone | 1.53 | 24 | Methyl-acetate | 0.49 |
| 11 | Propionic-acid | 1.37 | 25 | Methyl-furyl-cetone | 0.37 |
| 12 | Ethanal | 1.25 | 26 | Cronolactone | 0.37 |
| 13 | 2,3-butanedione | 1.25 | 27 | 2-methyl-2-cyclopenten-1-one | 0.37 |
| 14 | 2,3-pentanedione | 1.25 | 28 | Cyclopentenone | 0.37 |

Note: 28 model compounds were found in the Aspen Database from the original list of 34 compounds

*Table 5 Properties of biocrude sample from experimental results measured at standard conditions (4refinery-Scenarios for integration of bio-liquids in existing refinery processes, 2017) (Demirbas F. , 2010) (Pedersen, Jensen, Sandström, & Rosendahl, 2017)*

| | HHV(daf) [MJ/kg] | $\Delta h_f$ [MJ/kg]* | C [wt.%] (daf) | H [wt.%] (daf) | O [wt.%] (daf) | N [wt.%] (daf) | Density [kg/m3] |
|---|---|---|---|---|---|---|---|
| HTL biocrude | 35.90 | -2.22 | 80.00 | 8.40 | 11.00 | 0.50 | 1050.67 |
| Pyrolysis oil | 34.30 | -0.59 | 69.30 | 8.60 | 21.40 | 0.70 | 1195.00 |
| HVF 1 (300-350 C) | 35.50 | -3.05 | 77.80 | 9.20 | 13.00 | NR** | 1409.26*** |
| HVF 2 (>350 C) | 35.20 | -0.29 | 81.00 | 6.30 | 12.70 | NR** | 1517.30*** |

*Estimated from reported HHV (eq.6), ** Not reported, *** Estimated from HHV (Demirbas & Al-Ghamdi, 2015)

For the HTL biocrude, additional constraints were provided to the multi-objective optimization based on the reported distillation curve of the HTL biocrude sample and the boiling points of the model compounds: the sum of the mass fractions of model compounds whose boiling point is less or equal to the true boiling point of the cut (≤TBP) should not exceed the accumulated distilled fraction until that cut (eq. 22-23 for the first two cuts). In this way, the mass distribution of model compounds and the heavy fraction is intended to match bulk thermal and physical properties while resembling the distillation profile. The distillation profile data of the HTL biocrude is reported in Table 6.

$$\forall x_i: BP_i \leq 100: \quad \sum_{i=1}^{n} x_i \leq 1.6 \tag{22}$$

$$\forall x_i: BP_i \leq 150: \quad \sum_{i=1}^{n} x_i \leq 5.7 \tag{23}$$

$$\vdots$$

*Table 6 True boiling point and percentage of distilled fraction of HTL biocrude (Pedersen, Jensen, Sandström, & Rosendahl, 2017)*

| TBP [ºC] | Distilled fraction [wt. %] | Accumulated distilled fraction [wt. %] |
|---:|---:|---:|
| <100 | 1.6 | 1.6 |
| 100-150 | 4.1 | 5.7 |
| 150-200 | 6.3 | 12.0 |
| 200-250 | 6.0 | 18.0 |
| 250-300 | 15.3 | 33.3 |
| 300-350 | 10.3 | 43.6 |
| >350 | 51.8 | 95.4 |

In the study two optimization cases were evaluated: first giving all properties the same weight and second with the heat of formation as hard constraint. Additionally, there are more variables than equations in the system for which the solution depends on the initial point. Therefore, the optimized composition in the model is set as the average of the solutions generated from different initial random points which was arbitrarily set to 30. The results obtained from the optimized composition are compared with the experimental values and the estimated from the initial composition in Table 3 and Table 4 without optimization.

## 2.3 *Impact of models in energy balance of HTL*

A case study was used to assess the impact of the previous methodologies on the energy requirements. HTL was selected due to the high availability of data in-house and the main data used is summarized in Table 7. The wood characterization and the composition of the gas phase obtained were provided in the experiment while the composition of the aqueous phase at the outlet was adjusted to meet the reported TOC (~75 g/L) and minimize the errors in the elemental balance. The production of solids and water-soluble organics is not accounted for the yields reported in the 4refinery deliverable (4refinery-Scenarios for integration of bio-liquids in existing refinery processes, 2017).

*Table 7 Data reported on HTL experiment-4refinery biocrude (4refinery-Scenarios for integration of bio-liquids in existing refinery processes, 2017)*

| | | |
|---|---:|---:|
| T [ºC] / P [bar] | 400/300 | |
| Mass ratio at inlet (wood: aq. phase) | 1:3.2 | |
| | Mass yield | HHV (daf) [MJ/kg] |
| Biomass | -- | 19.56 |
| Biocrude | 0.45 | 35.90 |
| Gas | 0.45 | 7.70 |
| Water | 0.1 | -- |

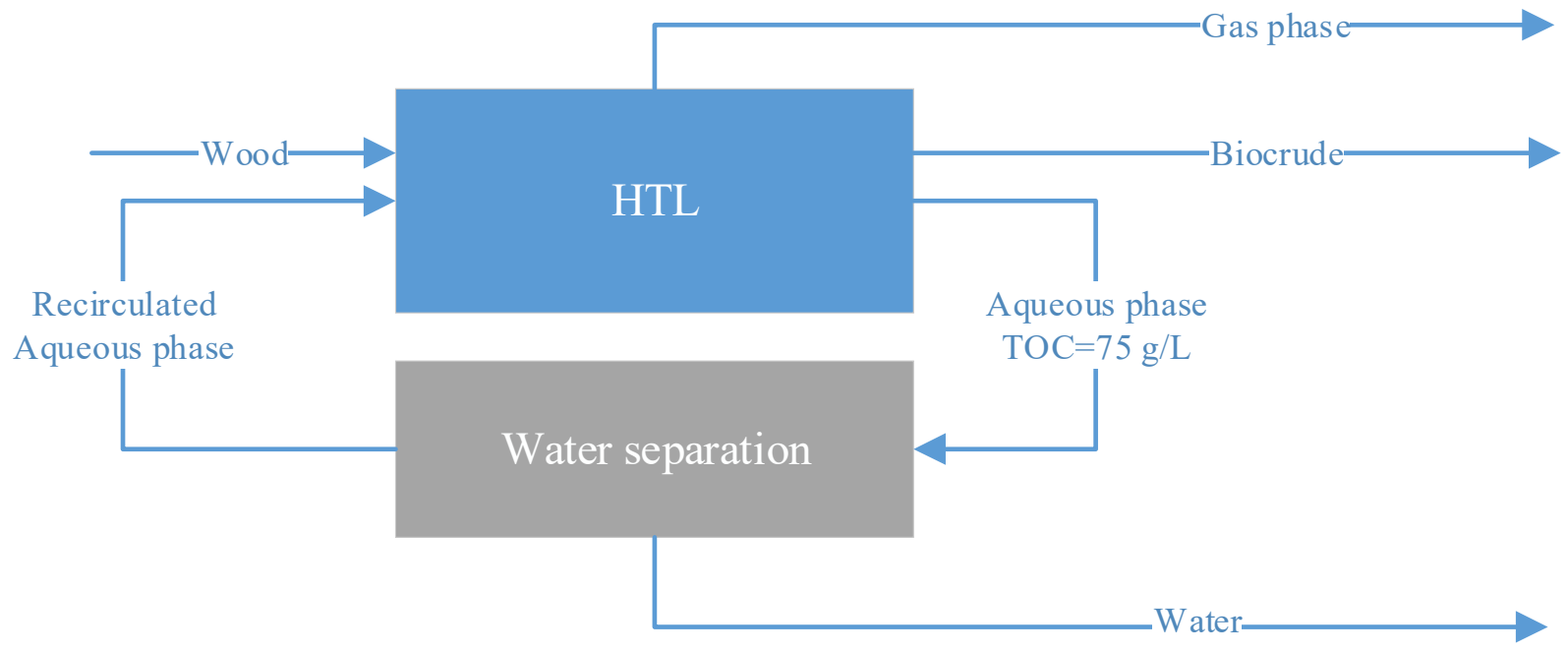

*Figure 4 Simplified process diagram of HTL experiment-4Refinery deliverable CBS1-369*

The case study was simulated in Aspen Plus® due to its extensive compound library. In total, 5 simulation cases were evaluated to assess the impact of property methods. The property method selected for the model compounds was SRK suitable to handle conditions close to the critical point of water. Wood was defined as a non-conventional component (solid component that cannot be characterized by a molecular formula) while the other phases (biocrude, gas and aqueous phase) are modeled through conventional components selected from the Aspen Plus® database in addition to the heavy fractions defined as pseudocomponents. In the case of the wood, the enthalpy model selected was HCOALGEN (general coal enthalpy model). This model includes different calculation options for the heat of combustion, heat of formation and specific heat capacity that were modified to evaluate the simulation cases of interest (Table 8). The HTL reactor was modeled as a yield reactor (RYield), in which the yields of the products are specified. Heating and cooling of the reactants and products was also simulated from standard to reaction conditions and vice versa.

*Table 8 Simulation cases in Aspen Plus®*

| | Initial | $\Delta h_f$ wood | Cp wood | Biocrude model | Overall adjusted |
|---|---|---|---|---|---|
| Wood | BOIE, Kirov correlation (Default Aspen Plus® HCOALGEN) | $\Delta h_f$ wood, Kirov correlation | BOIE, Cp for wood | BOIE, Kirov correlation | $\Delta h_f$ wood, Cp for wood |
| Biocrude | Not optimized | Not optimized | Not optimized | Optimized | Optimized |

In the initial case the default settings of HCOALGEN are used to model the wood: the heat of combustion (HHV) is estimated by Boie correlation, the heat of formation is estimated from the HHV, and the heat capacity is estimated using the Kirov correlation. The biocrude is modeled using the non-optimized composition of model compounds. The other cases evaluate one by one the developed methodologies for wood and biocrude (heat of formation of wood and optimized composition of biocrude). The impact of the specific heat capacity of the wood in the energy requirements was also evaluated by implementing a correlation for dry wood available in literature. For the biocrude, the heat capacity is estimated internally from the optimized composition of model compounds and used in the energy balance. Finally, an overall adjusted case was compared to the initial one.

## 3. RESULTS AND DISCUSSION

### 3.1 *Enthalpy of formation of woody biomass*

The fitted parameters of Lozano correlation (eq. 14) are presented in Table 9. The results of the HHV and heat of formation obtained with the correlations tested can be seen graphically in Figure 5a and 5b, with the upper and lower dashed lines representing the standard deviation. The numerical values are reported in Table 10.

*Table 9 Fitted parameters of Lozano correlation*

| | 1 | 2 | 3 | 4 | 5 | 6 | 7 | 8 | 9 | 10 | 11 |
|---|---|---|---|---|---|---|---|---|---|---|---|
| $a_i$ | -942.41 | -1067.07 | 459.78 | 154.56 | 173.56 | 8.59 | -5817.72 | -10.11 | -2.69 | 6.43 | 19689.26 |

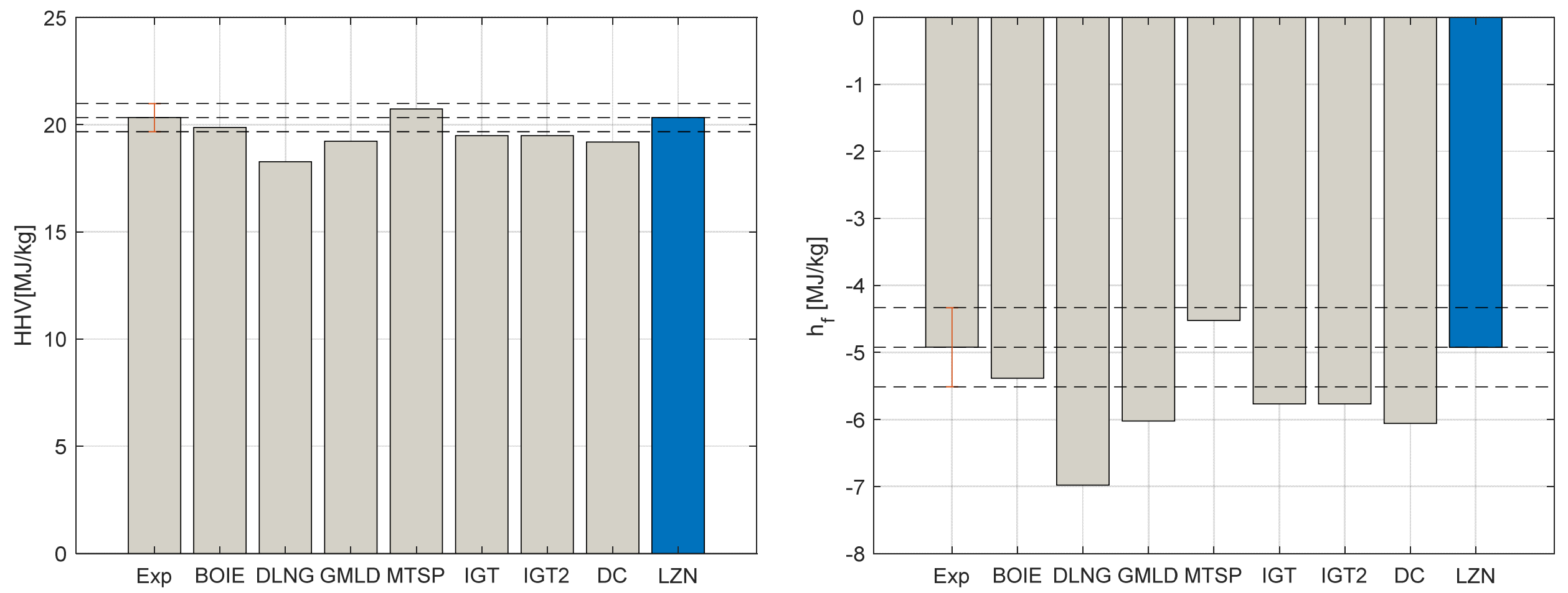

*Figure 5 (a) HHV and (b) Standard heat of formation from empirical correlations compared to experimental data*

*Table 10 Results of HHV, heat of formation and RMSE of empirical correlations for wood*

| [MJ/kg] | Exp | BOIE | DLNG | GMLD | MTSP | IGT | IGT2 | DC | LZN |
|---|---|---|---|---|---|---|---|---|---|
| HHV (daf) | 20.33 | 19.86 | 18.27 | 19.23 | 20.73 | 19.48 | 19.48 | 19.19 | 20.33 |
| Std [-] | 0.66 | -- | -- | -- | -- | -- | -- | -- | -- |
| $\Delta_f h$_std | -4.92 | -5.39 | -6.98 | -6.02 | -4.52 | -5.77 | -5.77 | -6.06 | -4.92 |
| Std [-] | 0.59 | -- | -- | -- | -- | -- | -- | -- | -- |
| RMSE | -- | 0.74 | 2.17 | 1.29 | 0.75 | 1.07 | 1.07 | 2.58 | 0.17 |
| RMSE_validation | -- | -- | -- | -- | -- | -- | -- | -- | 0.33 |

From these results it can be seen that, among the correlations tested, LZN (blue bar) has the best results for both properties. While for the HHV the difference is not outstanding compared to other correlations that give relative accurate results (BOIE, MTSP), it is more significant for the heat of formation in which the deviations of the existing correlations are significantly higher. This can be observed in Figure 6, in which the percentage error in the heat of formation is plotted vs the percentage error in the HHV for all correlations over the 30 samples tested. In most cases, an error in the HHV of ± 10% resulted in an error in the heat of formation between -60% to 40% approximately.

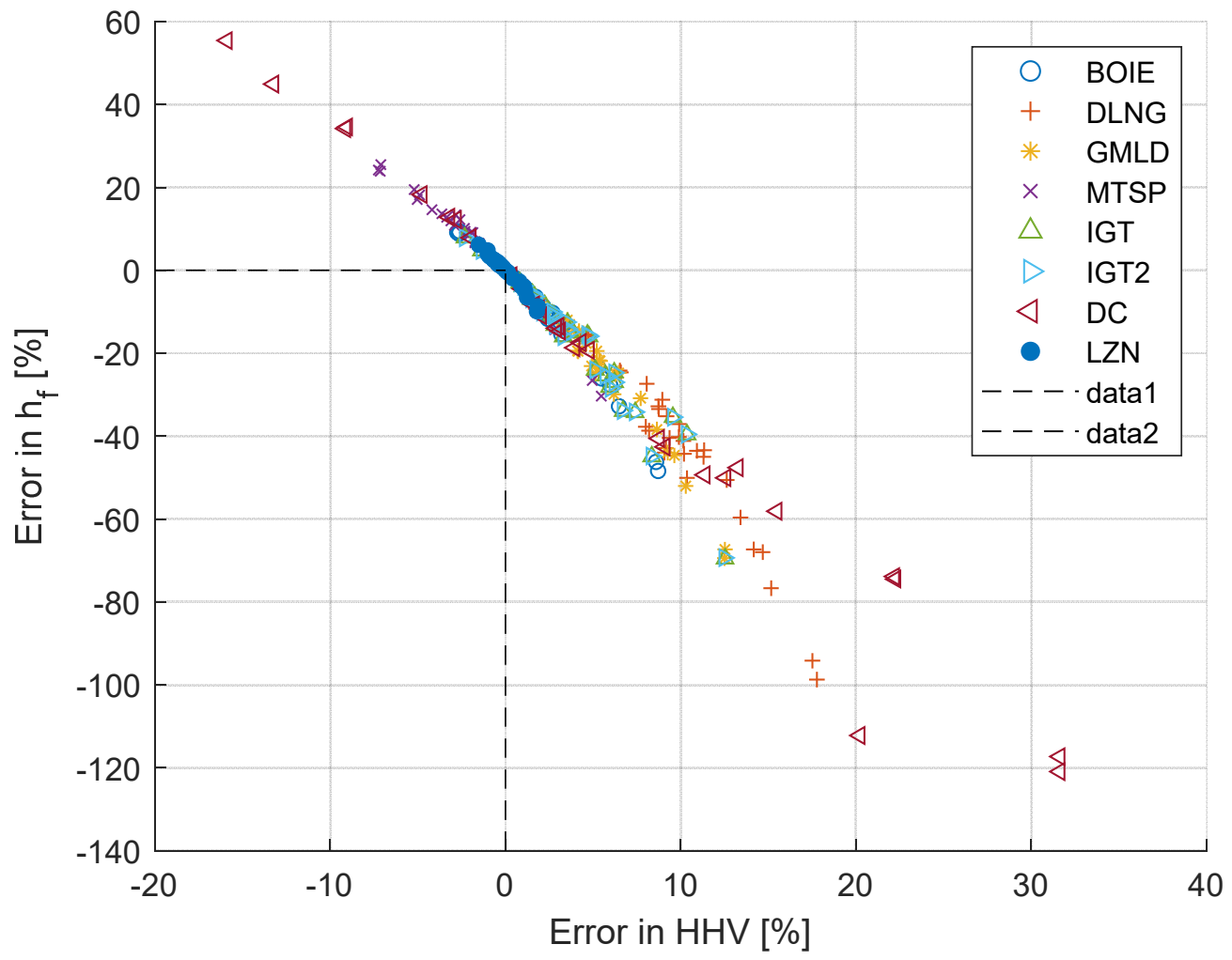

*Figure 6 Percentage error in HHV vs percentage error in heat of formation for tested correlations.*

Over 30 different samples, figure 7a shows that for LZN correlation there is a very good agreement between observed and predicted values, resulting in a RMSE of 0.17 MJ/kg. The figure also shows that the default options in Aspen Plus® should not be used uncritically, as this can introduce large deviations. Figure 7b shows the results of the cross validation with a RMSE of just 0.33 MJ/kg, still lower than the obtained for the existing correlations in which the minimum obtained was 0.74 MJ/kg.

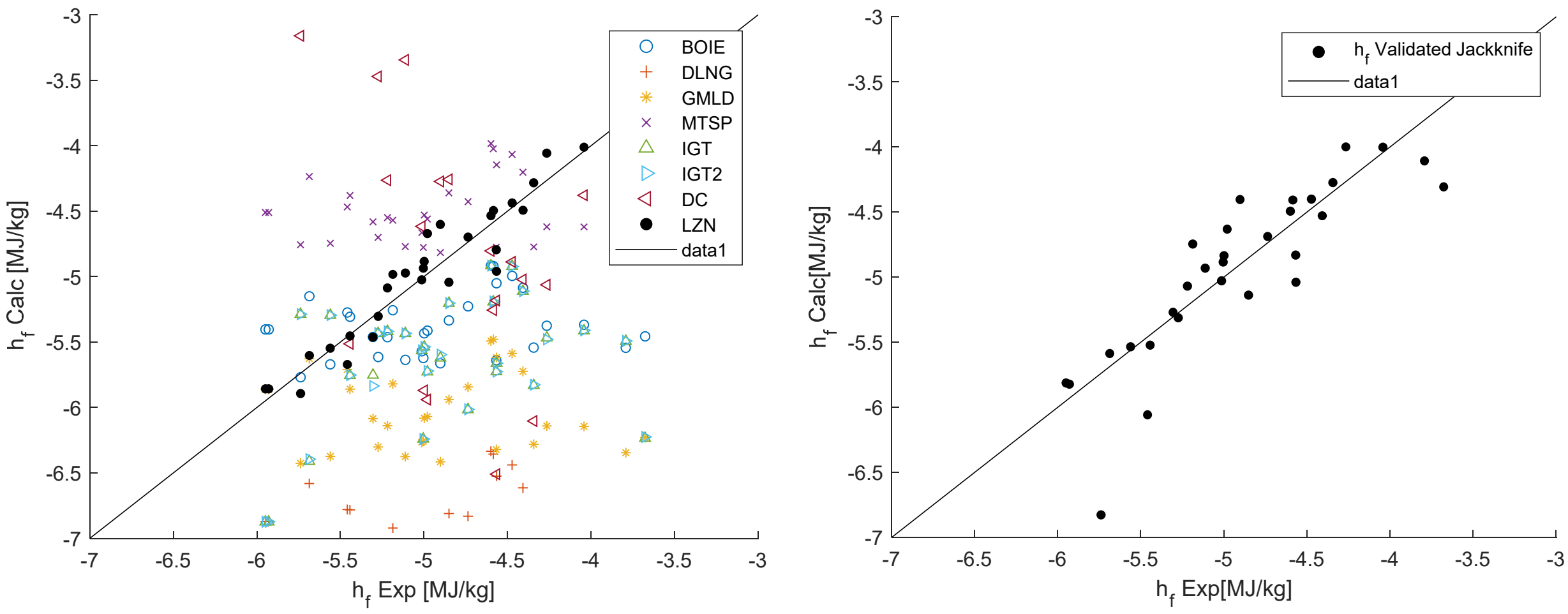

*Figure 7 Observed vs predicted $h_f$ for: (a) training dataset (b) Jackknife validation of LZN*

The previous results show that LZN correlation is significantly more accurate for estimating the heat of formation compared to the HHV route using the traditional correlations, in which small deviations in the estimation of the HHV have a very large impact.

## 3.2 *Optimized biocrude oils composition and properties*

This section discusses the results of the biocrude model regarding the mass distribution of model compounds and the thermochemical and physical properties analyzed for two cases: HTL biocrude and pyrolysis oil.

For the HTL biocrude, the results for the model compounds are presented next. In the figures on top (Figure 8a and 8b) all properties had the same weight, and in Figure 8c and 8d the heat of formation was introduced as hard constraint.

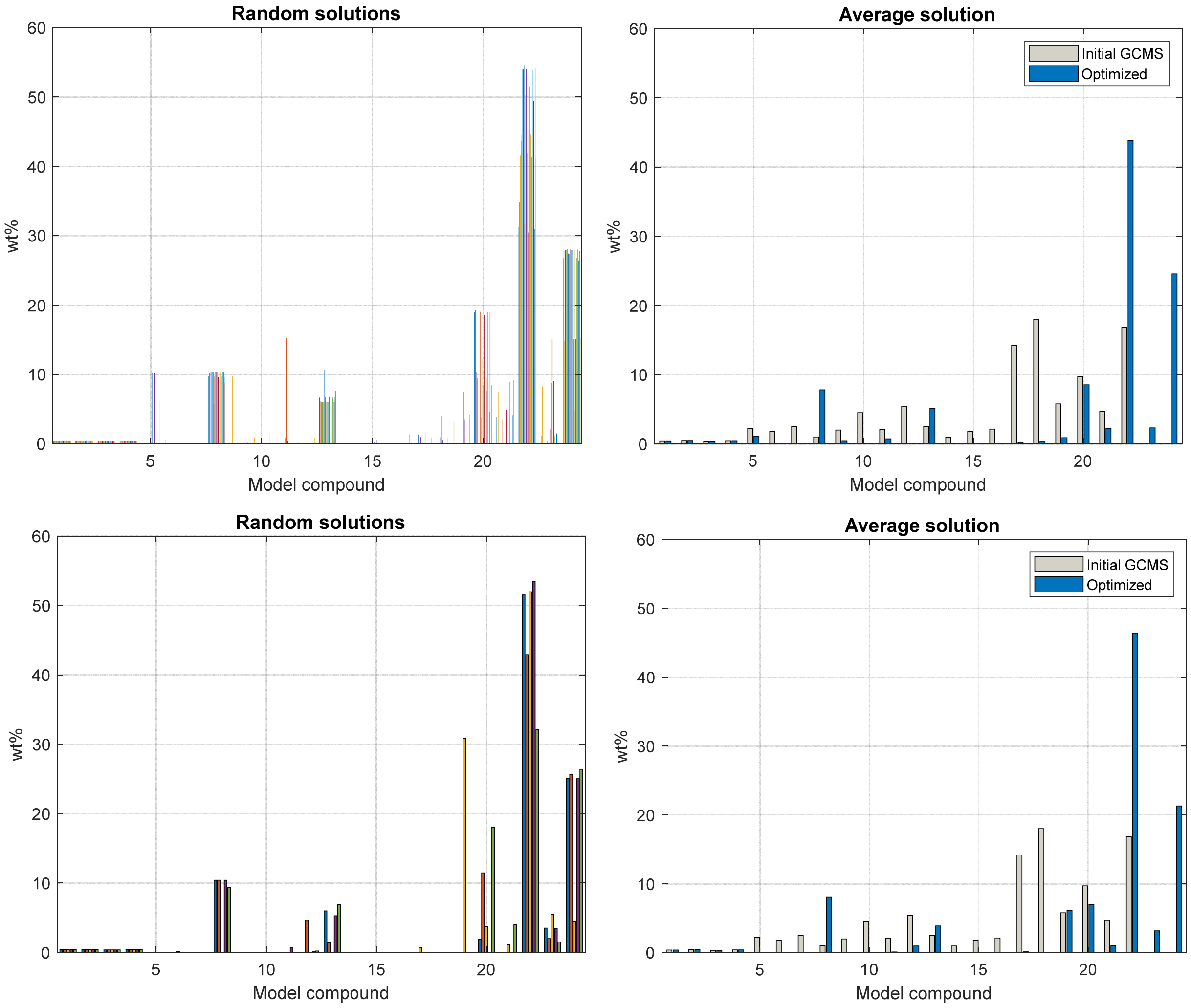

*Figure 8 Optimized weights of model compounds of HTL biocrude (a) and (b): All properties with same weight, (c) and (d): Heat of formation as hard constraint*

In general, for both cases it can be seen that some compounds are discarded while others are more consistently selected in all solutions (Figures 8a and 8c). When the heat of formation is introduced as hard constraint, the solution tends to increase the composition of the compound whose heat of formation is the closest to the objective value. However, in this case the error in the oxygen content is significantly increased while for the other properties is relatively constant, as can be seen in Figures 9a and 9b and Table 11.

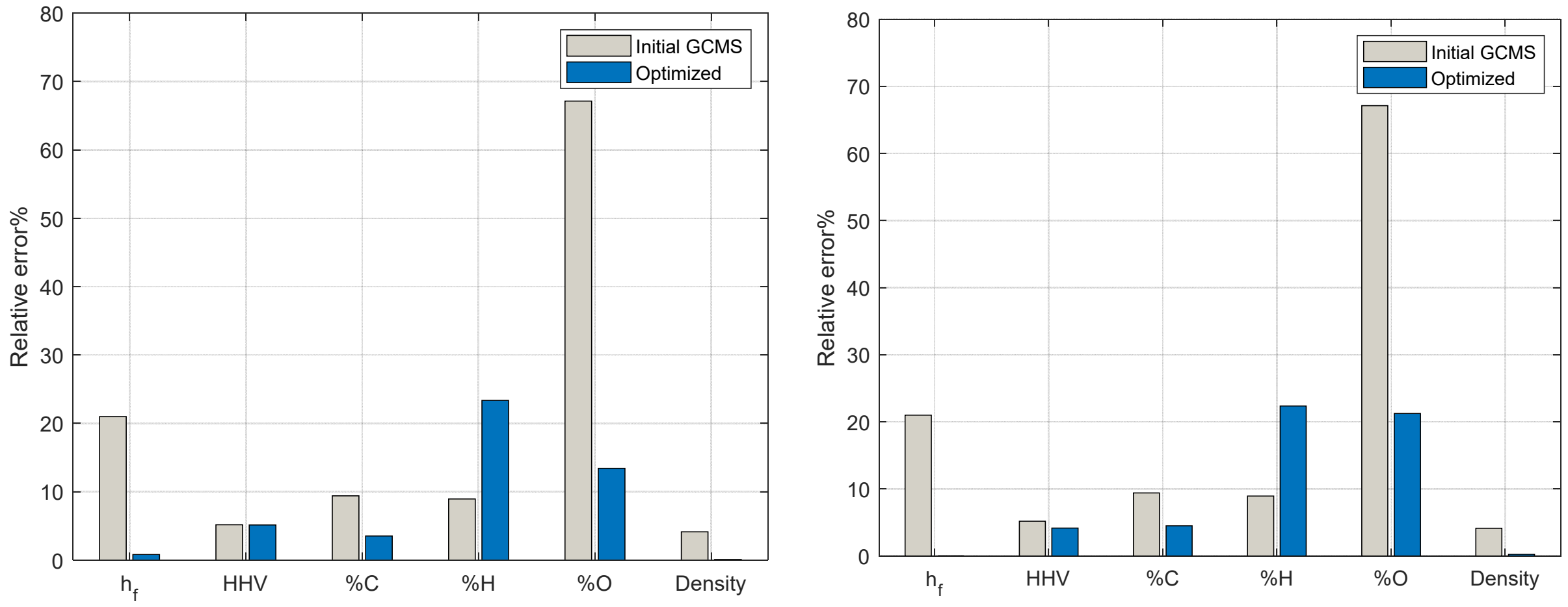

*Figure 9 Relative error in HTL biocrude properties based on initial GC-MS and optimized compositions for (a) same weights, (b) heat of formation as hard constraint*

*Table 11 Selected physical and thermal properties and elemental composition of HTL biocrude. Experimental and optimized results*

| | $\Delta h_f$ [MJ/kg] | HHV [MJ/kg] | C [wt.%] (daf) | H [wt.%] (daf) | O [wt.%] (daf) | Density [kg/m3] | Sum x |
|---|---|---|---|---|---|---|---|
| Reference | -2.22 | 35.90 | 80.00 | 8.40 | 11.00 | 1050.67 | -- |
| Initial values | -2.69 | 34.04 | 72.46 | 9.15 | 18.38 | 1007.19 | 100.00 |
| Relative error [%] | 20.99 | 5.18 | 9.42 | 8.95 | 67.12 | 4.14 | |
| Optimized (equal weights) | -2.24 | 37.75 | 77.16 | 10.36 | 12.48 | 1051.63 | 100.00 |
| Relative error [%] | 0.83 | 5.14 | 3.55 | 23.36 | 13.41 | 0.09 | |
| Optimized ($\Delta h_f$ hard constraint) | -2.22 | 37.39 | 76.39 | 10.28 | 13.34 | 1047.83 | 100.00 |
| Relative error [%] | 0.00 | 4.15 | 4.52 | 22.37 | 21.23 | 0.27 | |

These results show that the relative errors for most properties evaluated decreased significantly for the optimized compositions compared to the initial from the GC-MS peak areas. In the case of the HHV, the error did not decrease due to a higher error in the hydrogen content that counteract the lower error in the carbon. However, the error in the heat of formation, carbon and oxygen content and density was successfully minimized. By comparing the two optimization cases, the solution with all properties having the same weight is preferred as it gives better results in all the properties. The highest error is obtained for the hydrogen content. This can be explained by looking at the properties of the individual model compounds in Figure 10: the model compound with the highest hydrogen content (22. octadecanoic acid) is the one with the highest composition in the optimized solution (43.8%), while the composition of other model compounds with lower hydrogen content is not high enough to compensate. The fact that this compound is the most abundant regardless its high hydrogen content is explained by the constraints supplied to the model to match the distillation profile. In Figure 11a it can be seen that this compound together with the heaviest fraction (HVF2) are the only candidates that have boiling points above 350ºC, which is the most abundant fraction in the biocrude (Table 6).

The results of the distillation profile are presented in Figure 11b. It can be seen that for the optimized composition the profile obtained is much closer to the expected from experiments compared to the obtained from the initial GC-MS composition. Without optimization, the temperature profile has very large deviation and the values below 250ºC for all vaporized fractions evidence the lack of the heavy fractions. The largest deviations of the optimized profile are observed for the light cuts, indicating that other model compounds in the range of lower boiling points might be included in the analysis for better results. However, a good accuracy is observed for higher cuts which represent a larger fraction of the biocrude.

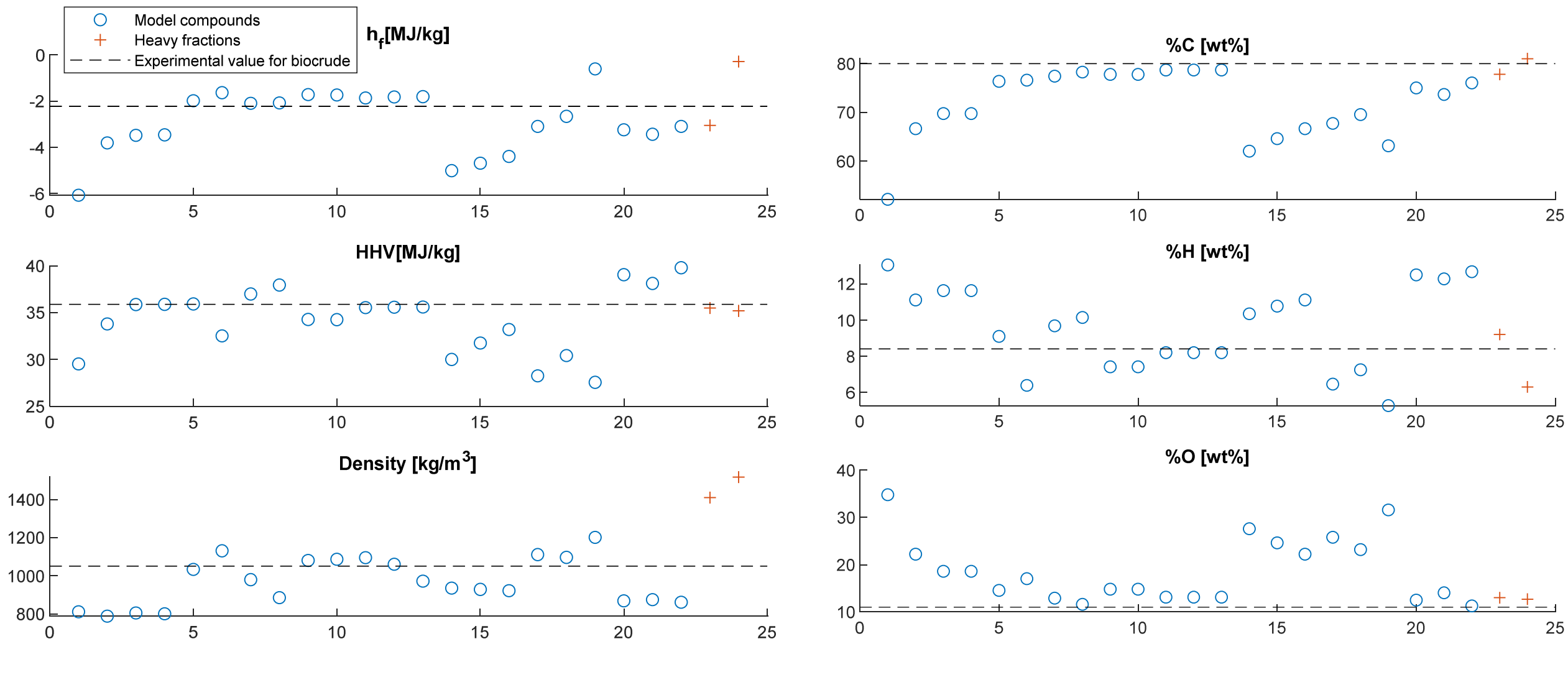

*Figure 10 Properties of individual model compounds compared with expected value of HTL biocrude at standard conditions. Blue symbols correspond to model compounds and red to HVF.*

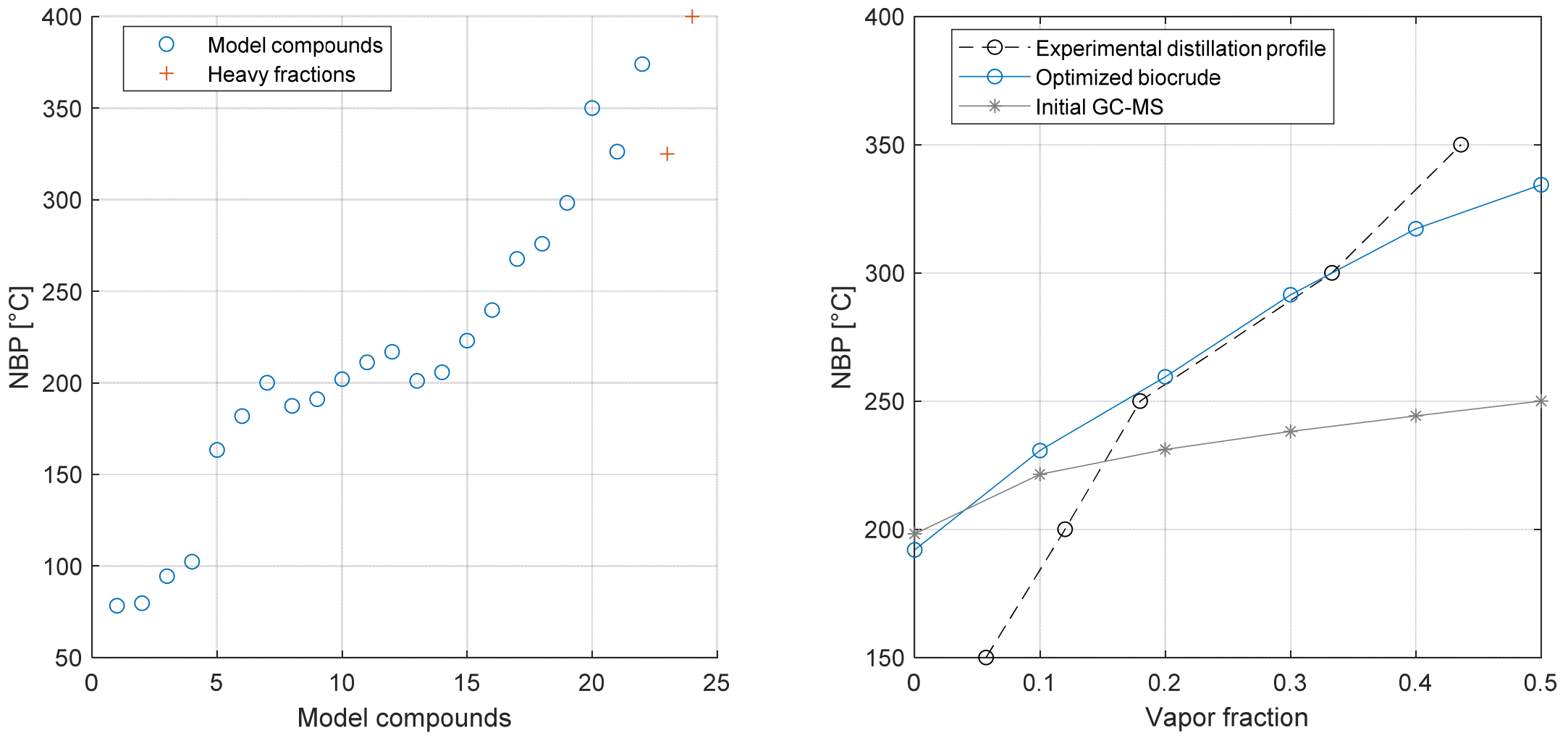

*Figure 11 (a) Normal boiling points of individual model compounds and heavy fractions (HVF1:average of the interval, HVF2: 400°C as indicative value >350°C) (b) Distillation profile of optimized and initial biocrude compositions compared to experimentally measured (Pedersen, Jensen, Sandström, & Rosendahl, 2017)*

*Pyrolysis Oil:*

For the pyrolysis oil the model was applied with all properties having the same weight. The results of the optimization are presented in Figure 12 and Table 12. It can be seen that, in general, the error was decreased even though the deviations are higher than in the previous case. This can be explained by looking at the properties of the individual model compounds (Figure 13), which show that for most model compounds the values are not in a good range to satisfy the objective functions, especially for the heat of formation and the oxygen content that had the highest deviations. Even though the properties of the heavy fractions give a better balance, the impact of these is lower than in the previous case as the distillation profile was not included in the optimization. The random solutions and average solution are shown in Figure 14.

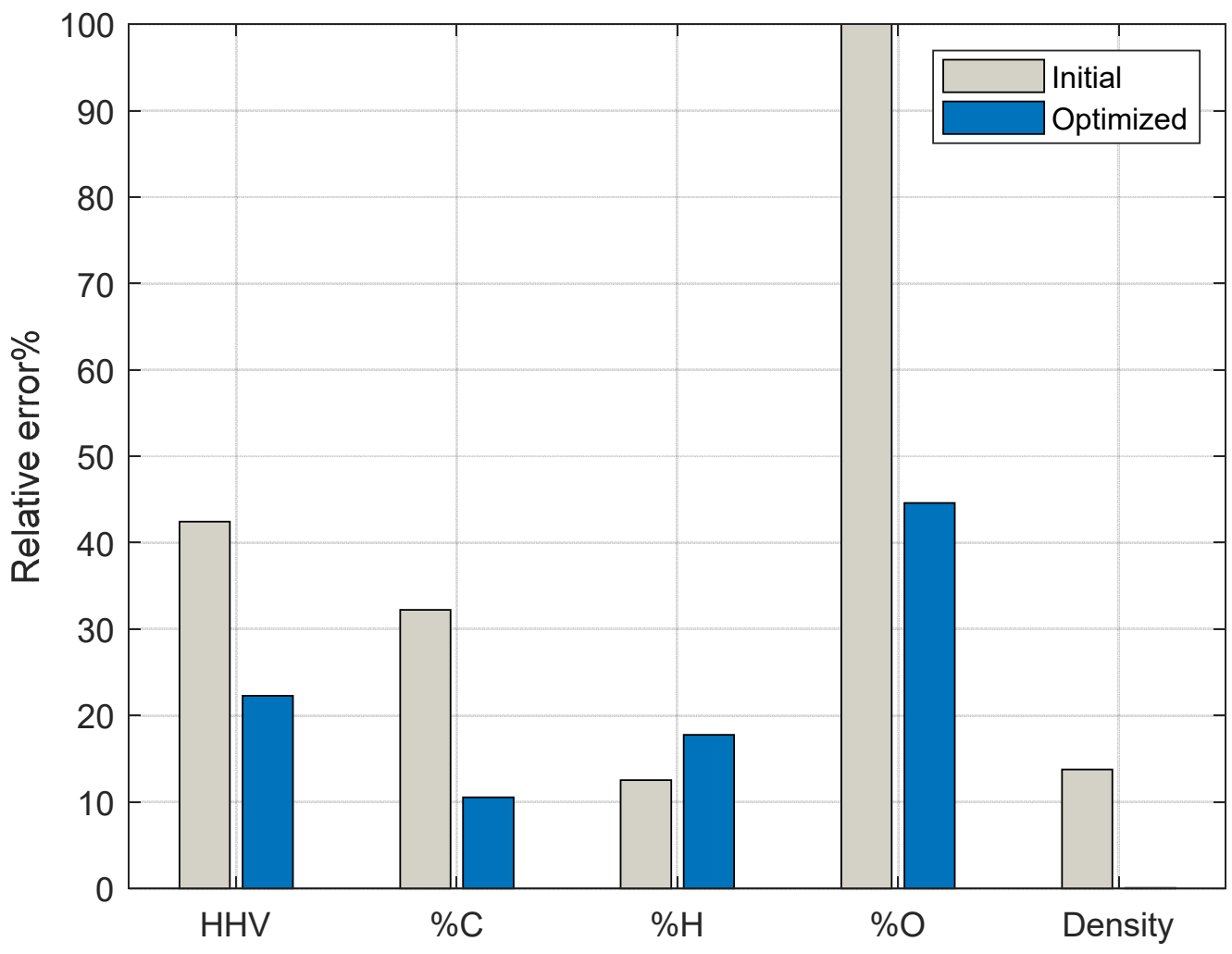

*Figure 12 Relative error in pyrolysis oil properties based on initial and optimized compositions*

*Table 12 Selected physical and thermal properties and elemental composition of pyrolysis oil. Experimental and optimized results*

| | $\Delta h_f$ [MJ/kg] | HHV [MJ/kg] | C [wt.%] (daf) | H [wt.%] (daf) | O [wt.%] (daf) | Density [kg/m3] | Sum x |
|---|---|---|---|---|---|---|---|
| Reference | -0.59 | 34.30 | 69.30 | 8.60 | 21.40 | 1195.00 | -- |
| Initial values | -6.16 | 20.19 | 47.95 | 7.50 | 44.55 | 1035.05 | 100.00 |
| Relative error [%] | 936.33 | 41.13 | 30.80 | 12.80 | 108.17 | 13.39 | 0.00 |
| Optimized | -3.69 | 26.66 | 61.98 | 7.07 | 30.94 | 1195.90 | 100.00 |
| Relative error [%] | 520.68 | 22.28 | 10.56 | 17.76 | 44.60 | 0.08 | 0.00 |

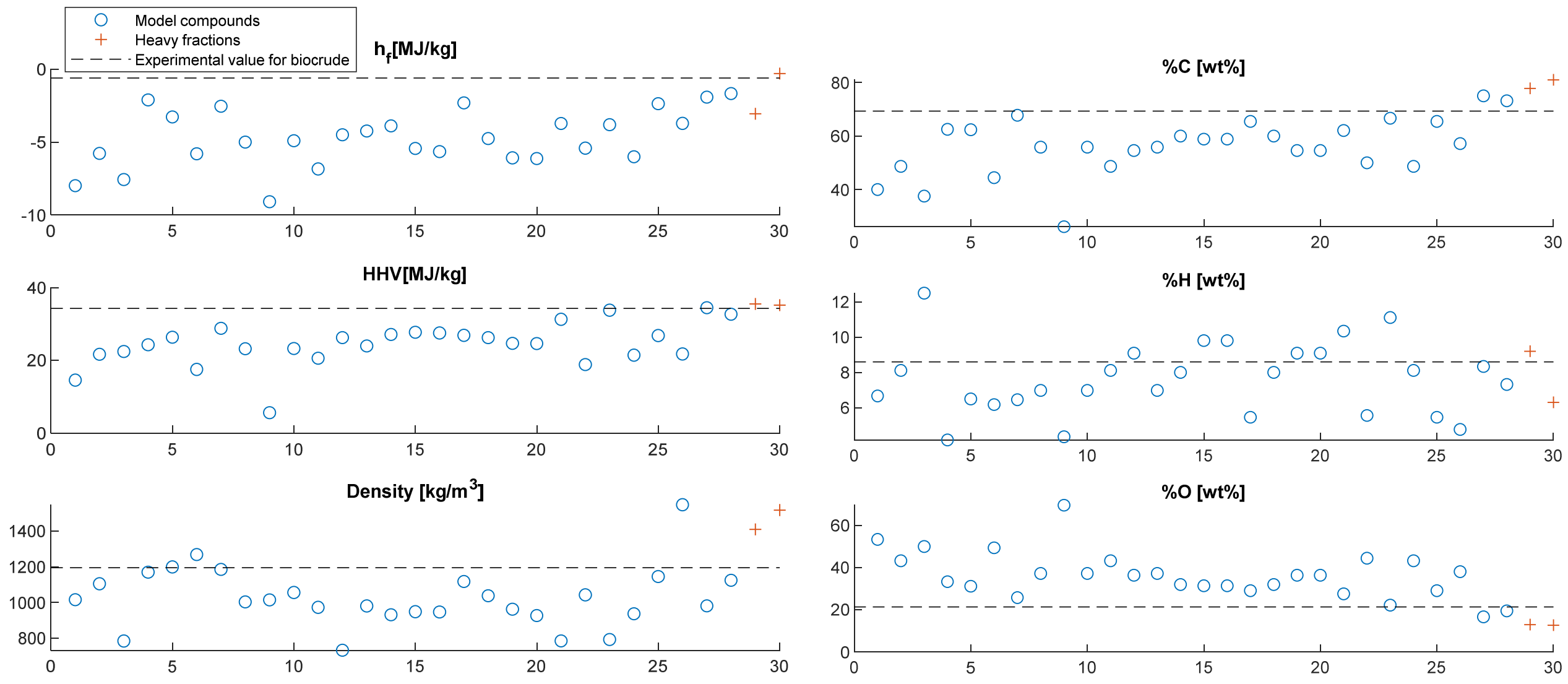

*Figure 13 Properties of individual model compounds compared with expected value of pyrolysis oil at standard conditions*

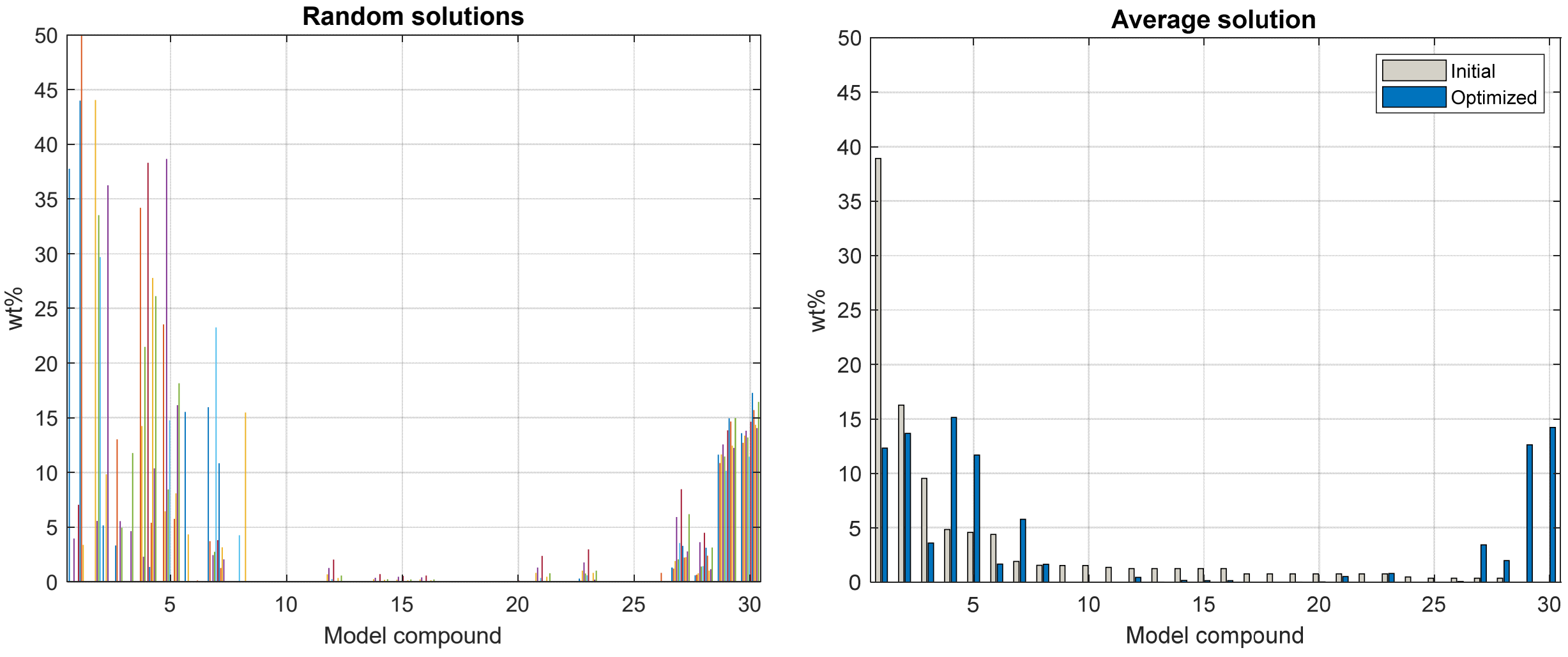

*Figure 14 Optimized weights of model compounds of pyrolysis oil. All properties with same weight*

## 3.3 *Impact of models in energy balance of HTL*

The results of the simulation cases developed in Aspen Plus ® are discussed in this section regarding the impact of the models developed in the reactor duty and the heating and cooling requirements. The results of the simulation cases are presented in Table 13. The experimental enthalpies of formation of wood and biocrude are included for comparison.

*Table 13 Results of simulation in Aspen Plus®*

| | Simulation cases | $\Delta h_f$ wood [MJ/kg] | $\Delta h_f$ biocrude [MJ/kg] | Duty HTL [MJ/kg wood] | Duty heater [MJ/kg wood] | Duty cooler [MJ/kg wood] |
|---|---|---|---|---|---|---|
| | Experimental | -5.07 | -2.22 | -- | -- | -- |
| 1 | Initial (Default HCOALGEN) | -5.60 | -2.69 | 0.29 | 6.69 | -8.06 |
| 2 | $\Delta h_f$ wood | -5.05 | -2.69 | -0.25 | 6.69 | -8.06 |
| 3 | Cp wood | -5.60 | -2.69 | 0.43 | 6.55 | -8.06 |
| 4 | BC model | -5.60 | -2.22 | 0.51 | 6.69 | -8.08 |
| 5 | Overall adj. | -5.05 | -2.22 | 0.12 | 6.54 | -8.08 |

Figure 15 shows the duties of the HTL reactor, the heater and the cooler for the different cases (negative values indicate energy release). It can be seen that the HTL duty changes significantly in all the cases analyzed compared to the duties of the heater and cooler, which are roughly constant and at least one order of magnitude higher. This indicates that the properties of the wood and the biocrude have a high influence on the reactor duty but do not influence significantly the requirements for heating and cooling the reactor streams, which is explained by the high proportion of water in the mixtures.

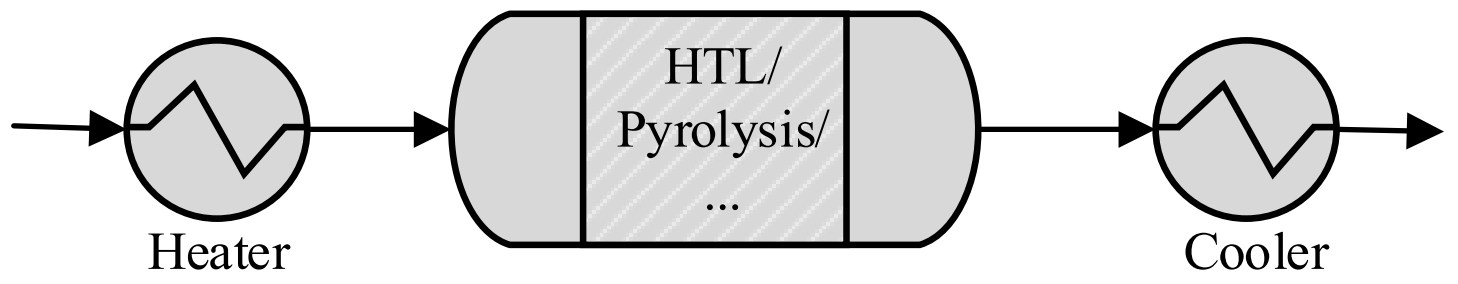

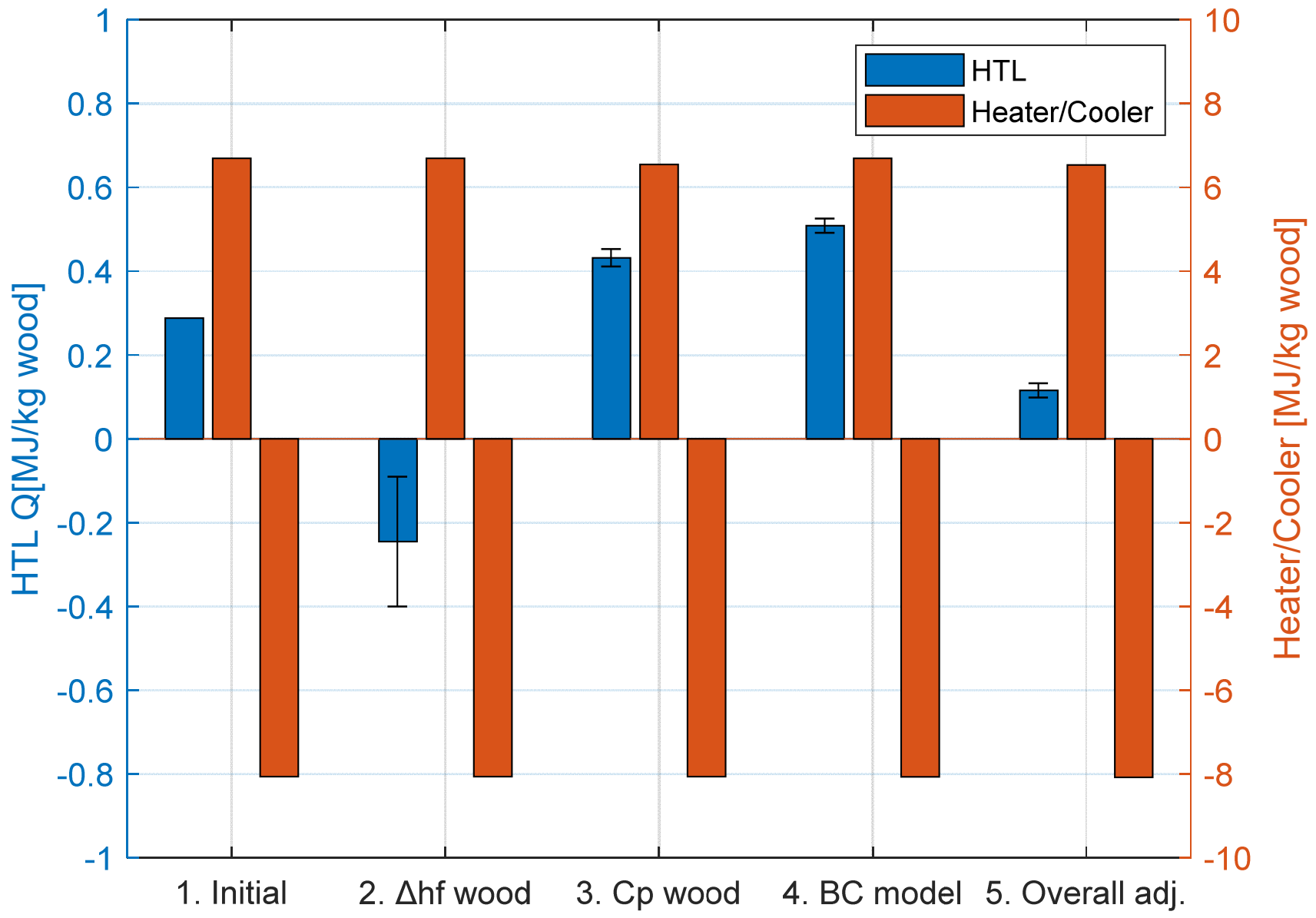

*Figure 15 Duty of HTL reactor and cooler/heater for different simulation cases*

Regarding the reactor duty, the result of case 1 (Initial- HCOALGEN) is compared to cases 2, 3 and 4 in which one property is adjusted at a time (e.g heat of formation of wood, heat capacity of wood and biocrude properties), and with case 5 in which all the adjustments are combined. In case 1 the enthalpies of both wood and biocrude are underestimated by 0.53 MJ/kg (10.4%) and 0.47 MJ/kg (21.4%) respectively.

In case 2, when the heat of formation of wood was adjusted, there was a dramatic change in the HTL duty that went from 0.29 to -0.25 MJ/kg wood, going from endothermic to exothermic. This is due to the underestimation of the heat of formation of wood in the initial case that leads to an overestimation of the reactor duty. The error bars show that the reactor duty changes more than ±50% when the heat of formation deviates ± 0.17 MJ/kg (3%), which corresponds to the RMSE of LZN correlation. It was estimated that in order to have an error in the reactor duty within the 10%, the error in the heat of formation should be around 0.4%. This shows that the heat of formation of wood has a major impact in the energy balance.

In case 3, when the heat capacity was corrected for wood (decrease of ~25%), the HTL duty increased 50% approximately. Different correlations reported in literature for dry wood show that the heat capacity varies between 1.2 and 2.0 kJ/kgK in temperatures between 20-200ºC (Hankalin, 2009) (Basu, 2010) (Grønli, 1996) (TenWolde A, 1988), which is around 0.5 kJ/kg lower than the predicted by the Kirov correlation (Figure 16). Lower values of heat capacity in the reactants induce higher values of Q, as can be deducted from eq. 1 and eq. 2. Above 200ºC, it seems that the heat capacity either continues to increase (Wildes correlation) or tends to a value around 2 kJ/kgK (Grønli correlation). .240790els3.The error bars in Figure 15 for case 3 represent these two behaviors and show that this variation is very small compared to the previous case.

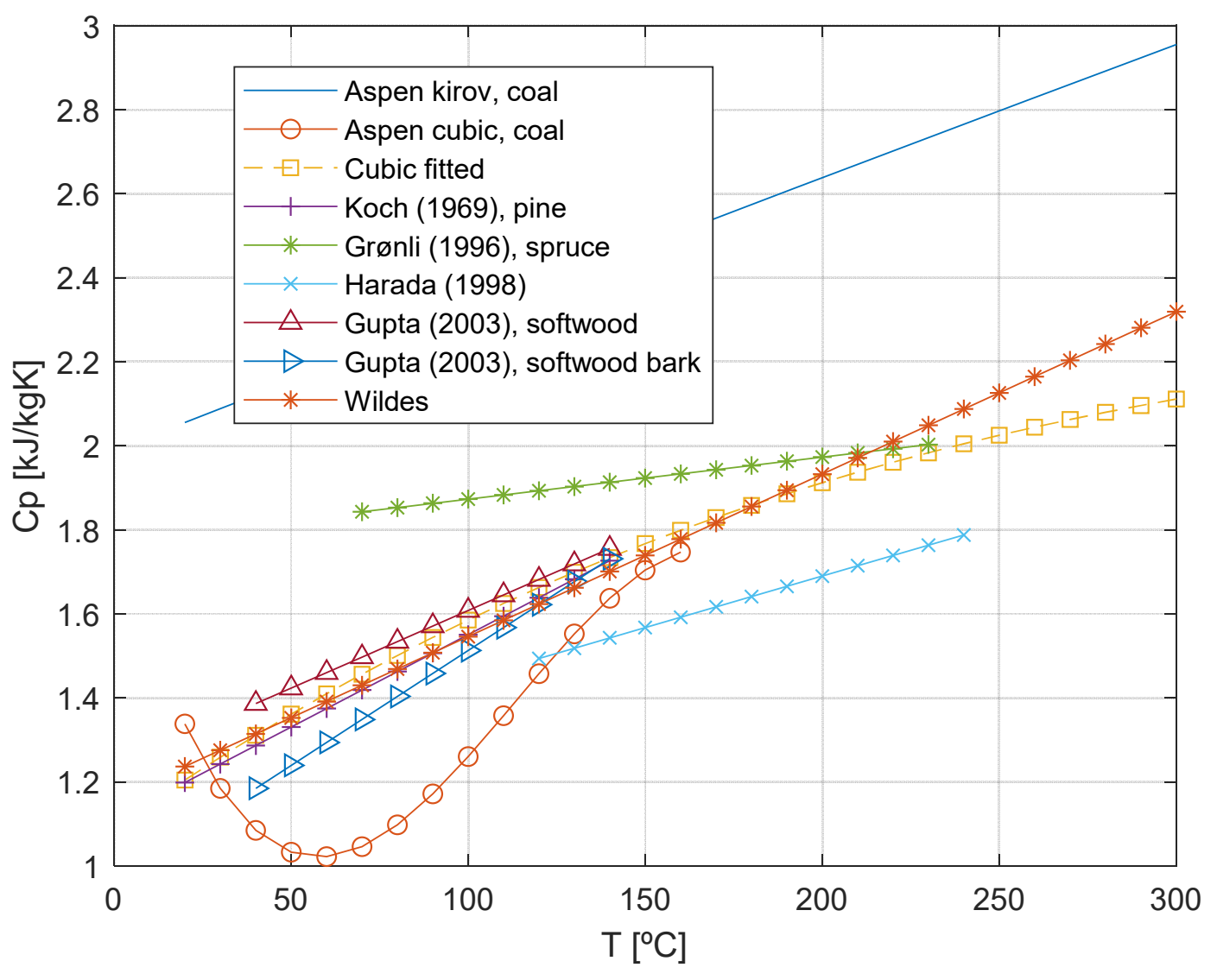

*Figure 16 Heat capacity of wood from different correlations*

In case 4, when the biocrude model was implemented, the heat of formation of biocrude increased by 17.5% (+0.47 MJ/kg) resulting in an increase of 77% approximately in the reactor duty. The error bars represent the two optimization cases –equal weights and heat of formation as hard constraint-, showing that the impact of this variation is low compared to the overall change introduced by the model.

Finally, in case 5, the adjustments in the properties of both wood and biocrude are integrated resulting in a reactor duty of 0.12 MJ/kg of wood, which is approximately 50% of the initial value, when no adjustment is implemented.

The results of the case study show that the reactor duty was highly sensitive to alterations in all the variables, changing in all cases by more than 50% with fluctuations in the properties between 5%-25% approximately. However, the variable with the highest impact was the heat of formation of wood, followed by the biocrude and the heat capacity of the wood. This result can be explained by looking at the enthalpies of reactants and products (Figure 17). The energy content of the wood is larger compared to the biocrude and there is less biocrude and gas produced than wood being processed. Additionally, since in the case study the aqueous phase had the same characteristics at the inlet and outlet of the reactor, this stream does not have an impact in the energy balance despite of its larger energy content, which overall gives more weight to the wood term in the heat balance (eq.1). It has to be taken into account that the yields are a determinant variable that indicates how the energy available in the system is being distributed among the products. The difference between blue and red bars in Figure 17 show the relatively small impact of the sensible heat in the enthalpy of the woody biomass according to (eq.2). Furthermore, it can be observed that the energy content of the gas phase produced is significant and could be potentially used for heating inside the process.

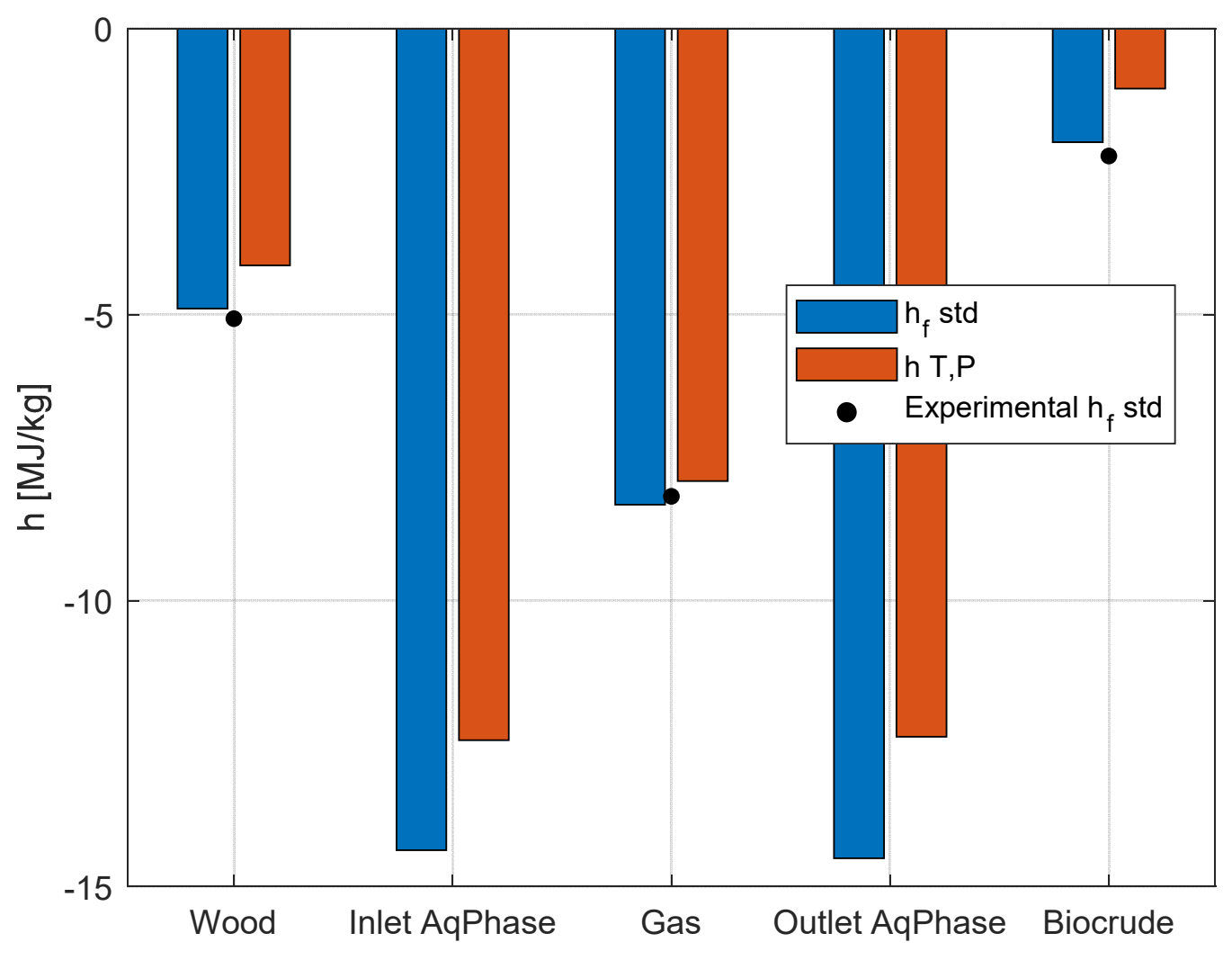

*Figure 17 Specific enthalpies of reactants and products obtained in Aspen Plus® vs values from experiments*

The results for the biomass and biocrude models show that, after optimization, the errors in the estimation of their thermal and physical properties are minimized and the standard enthalpies are close to the expected from experiments (black dots in Figure 17). From this, it is expected that the estimation of the reactor duty is much more reliable and closer to the actual value, providing a stronger basis for further process designs and heat integration analysis.

## 4. CONCLUSIONS

The modeling approach developed contributes to a more accurate representation of the properties of biomass and biocrude oils and provides a more reliable estimation of the reactor duty in the modeling of thermochemical processes in which the kinetics and stoichiometry are not established.

Over 30 different samples of woody biomass, the correlation developed in this work (Lozano correlation) was significantly more accurate for estimating the heat of formation compared to the HHV route using the traditional correlations. In most of the cases analyzed, the traditional correlations had an error in the HHV between 5-10% that result in under-estimation of the heat of formation by -20 to -60%, while for LZN correlation the error in the heat of formation was in all cases below 6.5%. These results are considered very satisfactory taking into account that the RMSE obtained was 0.33 MJ/kg compared to a standard deviation of the samples of 0.59 MJ/kg, and that the size of the sample for the parameter estimation was relatively small.

From the results of the biocrude model it can be concluded that this approach was successful in providing a more accurate representation of the physical and thermochemical properties. The errors in each property evaluated decreased significantly for the optimized composition compared to the initial from the GC-MS results and the distillation profile is significantly improved, which is crucial for further analysis of heat requirements in the upgrading process. The results show that, in general, having a larger number of model compounds is not necessarily translated into a more accurate representation of the biocrude. Further improvements of the model can be achieved by the inclusion of additional equations or constraints.

Regarding the case study, it was concluded that the properties of the wood and the biocrude had a high influence on the reactor duty but did not influence significantly the requirements for heating and cooling the reactor streams, which were roughly constant and at least one order of magnitude higher than the reactor duty. This was explained by the high proportion of water in the mixtures and the assumption of aqueous phase with the same composition at the inlet and outlet of the reactor. The variable that had the highest impact in the reactor duty was the woody biomass heat of formation, followed by the biocrude properties and the heat capacity of the wood, showing the importance of having an accurate estimation. Previous studies consulted that deal with the modeling of biomass conversion processes in Aspen Plus® use HCOALGEN without providing detailed information on the correlations selected for biomass, and/or the model compounds approach without optimization of the biocrude properties (Tzanetis, Posada, & Ramirez, 2017) (Pedersen, Hansen, Miralles, Villamar, & Rosendahl, 2017) (Adeyemi & Janajreh, 2015) (Stephen G. Gopaul, 2014) (W. Doherty, 2013). The results of the case study show that the correlations selected have a significant impact in the reactor duty, for which, these should be carefully selected to have more reliable results in the process design.

## 5. ACKNOWLEDGMENT

This project has received funding from the European Union's Horizon 2020 research and innovation program under grant no. 727531 (4refinery) and grant no. 765515 (Marie Skłodowska-Curie ITN, ENSYSTRA).